\documentclass[twocolumn,showpacs,preprintnumbers,amsmath,
               nofootinbib,aps]{revtex4}

\usepackage{graphicx}% Include figure files
\usepackage{dcolumn}% Align table columns on decimal point
\usepackage{bm}% bold math

\newcommand*{\no}{\noindent}
\newcommand*{\bea}{\begin{eqnarray}}
\newcommand*{\eea}{\end{eqnarray}}
\newcommand*{\be}{\begin{equation}}
\newcommand*{\ee}{\end{equation}}
\newcommand*{\pd}{\partial}

\newcommand*{\pref}[1]{(\ref{#1})}

\newcommand*{\mn}{{\mu\nu}}

\newcommand*{\nn}{\nonumber}

\begin{document}

\preprint{}

\title{More on Gribov copies and propagators in Landau-gauge Yang-Mills theory}

\author{Axel Maas}\email{axel.maas@uni-graz.at}
\affiliation{Department of Theoretical Physics, Institute of Physics,\\
             Karl-Franzens University Graz, Universit\"atsplatz 5, A-8010 Graz,
             Austria}
\affiliation{Department of Complex Physical Systems, Institute of Physics, \\
             Slovak Academy of Sciences, D\'{u}bravsk\'{a} cesta 9, SK-845 11 Bratislava,
             Slovakia}

\date{\today}

\begin{abstract}
Fixing a gauge in the non-perturbative domain of Yang-Mills theory is a non-trivial problem due to the presence of Gribov copies. In particular, there are different gauges in the non-perturbative regime which all correspond to the same definition of a gauge in the perturbative domain. Gauge-dependent correlation functions may differ in these gauges. Two such gauges are the minimal and absolute Landau gauge, both corresponding to the perturbative Landau gauge. These, and their numerical implementation, are described and presented in detail. Other choices will also be discussed.

This investigation is performed, using numerical lattice gauge theory calculations, by comparing the propagators of gluons and ghosts for the minimal Landau gauge and the absolute Landau gauge in SU(2) Yang-Mills theory. It is found that the propagators are different in the far infrared and even at energy scales of the order of half a GeV. In particular, also the finite-volume effects are modified. This is observed in two and three dimensions. Some remarks on the four-dimensional case are provided as well.
\end{abstract}

\maketitle

%%%%%%%%%%%%%%%%%%%%%%%%%%%%%%%%%%%%%%%%%%%%%%%%%%%%%%%%%%%%%%%%%%%%%%%%%%%%%%%%%%%%%%%%%%%%%%%%%%%

\section{Introduction}

Green's functions encode the full dynamics of a theory \cite{Rivers:1987hi}. In gauge theories these are in general gauge-dependent, and it is necessary to determine them in a fixed gauge. In the perturbative regime, this can be conveniently done by local gauge conditions \cite{Bohm:yx}. However, to determine the Green's functions over the full momentum range it is necessary to extend the definition of a gauge into the non-perturbative domain. Unfortunately, there is no unique local possibility to extend the local gauge-fixing conditions of perturbation theory into this regime. One is therefore faced with a, possibly innumerable, number of gauges all corresponding to the same perturbative gauge. This will be further discussed in section \ref{slandau} and \ref{ssgauge} for the example of Landau gauges. In particular, two different Landau gauges will be investigated here in detail, the minimal Landau gauge and the absolute Landau gauge.

The importance of differing non-perturbative completions of perturbative local gauge conditions is that they potentially lead to differences in the gauge-dependent Green's functions. Of course, these differences have to become subleading at sufficiently large momenta, where non-perturbative effects become sub-leading. However, and as will be shown in section \ref{sgauge}, these differences may be qualitative at low momenta, and may still be quantitatively relevant at intermediate momenta of the order of the scale $\Lambda_{\mathrm{YM}}$ and above. Hence, not only calculations in the asymptotic infrared momentum range have to take this into account. Precision calculations of observable quantities, which are based on gauge-dependent correlation functions, may also need to take care of this. Apart from the practical purpose of calculating the Green's functions, this problem is also of a more fundamental interest. Confinement scenarios, like the one of Gribov and Zwanziger \cite{Gribov,Zwanziger:2003cf,gzwanziger,gzwanziger2} and the one of Kugo and Ojima \cite{Kugo} (GZKO), give predictions for correlation functions in certain, well-defined non-perturbative gauges. To check these predictions and hence the scenarios, it is necessary to determine the Green's functions in exactly these gauges.

The determination of the Green's functions is not a trivial problem. To obtain them in the full momentum range, various methods have been employed. Continuum methods, like Dyson-Schwinger equations \cite{von Smekal:1997vx,Alkofer:2000wg,Watson:2001yv,Lerche:2002ep,Fischer:2002eq,Alkofer:2004it,Schleifenbaum:2004id,Fischer:2006ub,Fischer:2006vf,Alkofer:2008jy,Kellermann:2008iw,Huber:2007kc,Maas:2004se,Aguilar:2004sw,Binosi:2007pi,Aguilar:2007nf,Aguilar:2008xm,Boucaud:2006if,Boucaud:2008ji,Boucaud:2008ky,Bloch:2002eq,dse}, functional renormalization group equations \cite{Fischer:2006vf,Pawlowski:2003hq,Fischer:2004uk,Braun:2007bx}, stochastic quantization approaches \cite{Zwanziger:2003cf,gzwanziger} or effective action methods \cite{Dudal:2005na,Dudal:2007cw,Capri:2007ix,Dudal:2008sp,Dudal:2008rm} have to rely on assumptions, in general. Lattice gauge theory \cite{Oliveira:2007dy,4d,sternbeck,sternbeck06,Cucchieri:2006xi,cucchieril7,l72,Cucchieri:2007rg,Bogolubsky,limghost,gggvertex,Ilgenfritz:2006he,Cucchieri:2008qm,Cucchieri:2003di,Cucchieri:1997ns,Cucchieri:2007ta,Maas:2007uv,Cucchieri:2006tf} is not in need of such assumptions, but suffers from artifacts, in particular finite volume effects \cite{cucchieril7,l72,Cucchieri:2007rg,limghost,Cucchieri:2007ta,Fischer:2007pf}. On top of these problems comes the necessity to fix the same non-perturbative definition of a gauge in all methods. For functional calculations this can possibly be implemented by means of boundary conditions \cite{dse}. In case of lattice gauge theory, this is a hard numerical problem, which has been investigated previously \cite{sternbeck,Bogolubsky,Cucchieri:1997ns,mmp4,Silva:2004bv} following the pioneering work \cite{Cucchieri:1997dx}, and will be discussed in great detail here as well.

Therefore, neither approach alone has been able to provide a final result for the Green's functions. For a reliable comparison therefore all of these problems have to be addressed. However, the uncertainties due to the inherent problems in this comparison have sparked quite a prolonged debate on the infrared behavior of Green's functions, and whether they are in agreement with the predictions of the GZKO scenarios or not, for a summary see \cite{dse}. In particular, the investigation with lattice calculations has, so far, not yet provided a clear picture. This will be discussed in more detail in section \ref{sgauge}, and the issue of gauge fixing on the lattice is covered in section \ref{sgf} and appendix \ref{sabslan}. 

In addition to the problem of gauge-fixing, finite volume artifacts are the main problem \cite{cucchieril7,l72,Cucchieri:2007rg,limghost,Cucchieri:2007ta,Fischer:2007pf}. The influence of these artifacts has been investigated in detail \cite{cucchieril7,l72,Cucchieri:2007rg,limghost,Cucchieri:2007ta}, and seems to indicate a contradiction between the GZKO scenario and the actual form of the Green's functions\footnote{See, however, \cite{Dudal:2008sp,Dudal:2008rm} for a possible different interpretation.}. A possibility would be that the gauge in these investigations is not the same as in the continuum calculations. This will be discussed here. Unfortunately, the increase in complexity to fix alternative versions of the Landau gauge in the non-perturbative domain prevents the access of large lattice volumes at the current time in four dimensions. This discrepancy also exists in three dimensions \cite{cucchieril7,Cucchieri:2007rg,limghost}, while no discrepancy exists in two dimensions \cite{Cucchieri:2007rg,limghost,Maas:2007uv}. It is therefore useful to address the problem first in lower dimensions, where large volumes can be accessed more easily. There is another advantage when studying the problem first in lower dimensions: The fixing of the alternative gauges is easier in lower dimensions, as will be shown in section \ref{sgauge}. As a result, it will be found that the finite-volume effects between both gauges differ. It will also be discussed whether they may be the reason for the apparent discrepancy.

The organization of the presentation of this investigation is as follows: In section \ref{slandau}, general aspects of fixing Landau gauges in the non-perturbative domain are discussed. Some specific properties of gauge-fixing in connection with Gribov copies are discussed in section \ref{sgf}. Afterwards, the direct comparison of the propagators in the minimal and absolute Landau gauge in two and three dimensions is shown in section \ref{sgauge}. Some remarks on the four-dimensional case will also be made there. The results obtained are discussed in section \ref{sdiscuss}. In particular, a possible interpretation will be provided in subsection \ref{ssint}. Finally, everything is summed up in section \ref{ssum}. The algorithm to fix (or, at least, approach) the absolute Landau gauge employed here is presented in appendix \ref{sabslan}. Technical details of the lattice simulations performed can be found in appendix \ref{stech}.

\section{Landau gauges in the non-perturbative domain}\label{slandau}

In perturbation theory, Landau gauge is defined by the local condition
\be
\pd_\mu A_\mu^a=0\label{landau}
\ee
\no on the gauge field $A_\mu^a$. It belongs to the class of covariant gauges, which in general average over the (perturbative) gauge orbit with a given weight function \cite{Bohm:yx}. In case of the Landau gauge, this weight function is a $\delta$-function. Unfortunately, it has been found \cite{Gribov,Singer:dk} that this local condition is insufficient to select a unique representative of the gauge orbit: There are different elements of the gauge orbit, so-called Gribov-Singer copies, which all satisfy the condition \pref{landau}. An example is, e.\ g., the instanton \cite{Maas:2005qt}. The set of all elements of a gauge orbit satisfying condition \pref{landau} will be called the residual gauge orbit. It has been shown that there is no possibility to specify a unique element on the residual gauge orbit with any set of additional local conditions \cite{Singer:dk}.

Further specifications of the gauge will therefore necessarily be non-local. Any of these should reduce to the perturbative Landau gauge \pref{landau} in the perturbative limit. This implies that the residual global color symmetry may not be broken by the gauge fixing, except when this breaking is soft, i.\ e., vanishes in the perturbative limit. As, however, the Kugo-Ojima scenario \cite{Kugo} is based on an unbroken global color symmetry, the judicious choice is not to break it by the gauge condition.

It has been argued long ago that the gauge-fixing will be complete when the gauge field is further restricted to the first Gribov region \cite{Gribov}. This region is characterized by the semi-positiveness of the Faddeev-Popov operator, which in Landau gauge is given by
\be
-\pd_\mu(\delta^{ab}\pd_\mu-gf^{abc}A_{\mu}^c)\label{fpop}.
\ee
\no Unfortunately, this is insufficient \cite{vanBaal:1997gu}. Hence, it is necessary to further restrict the gauge field. A sufficient possibility\footnote{Actually, degenerate minima appear on the boundary of this region, which have to be identified \cite{vanBaal:1997gu}. This is resolved implicitly by the algorithm employed in the lattice calculations below by always choosing exactly one gauge copy for each configuration.} here is to restrict it to the fundamental modular region \cite{vanBaal:1997gu}, which is characterized by the condition that the gauge-fixed field configuration absolutely minimizes
\be
{\cal F}(A)=\frac{1}{2dV}\int d^dx A_\mu^a(x) A_\mu^a(x),\label{fumod}
\ee
\no where $d$ is the dimensionality and $V$ is the $d$-volume. It can further be shown that all minima of \pref{fumod} with \pref{fpop} positive lie in the first Gribov region \cite{gzwanziger}. Hence, the fundamental modular region is enclosed in the Gribov region. In fact, both have partly a common boundary \cite{gzwanziger2}, on which, e.\ g., instantons lie \cite{Maas:2005qt}. The condition \pref{fumod} can be rewritten after double Fourier transformation as
\be
{\cal F}(D)=\frac{N_g}{2^d\pi^{d/2}\Gamma\left(1+\frac{d}{2}\right)V}\int dp p^{d-1} D_{\mu\mu}^{aa}(p),\label{propmin}
\ee
\no where $D_\mn^{ab}$ is the gluon propagator and $N_g$ the number of gluons, 3 for SU(2). Therefore, this condition is equivalent to minimizing the total trace of the gluon propagator on the residual gauge orbit. Unfortunately, this is an extremalization problem in a high (in the continuum infinite) dimensional space. It is therefore NP-hard, similar to spin-glass problems. In particular, the complexity increases exponentially with lattice sites. This is one origin of the problems in gauge fixing in numerical lattice gauge theory calculations.

From these observations, it is possible to design different types of non-perturbative Landau gauges. One is the absolute Landau gauge, defined by finding the absolute minimum of \pref{fumod}. In the spirit of the perturbative Landau gauge, this prescription selects exactly one representative on the residual gauge orbit, the one in the fundamental modular region. A second possibility is to select an arbitrary representative of the residual gauge orbit, which lies inside the first Gribov region. This is the most commonly adopted choice in lattice calculations, and is termed the minimal Landau gauge, see e.\ g.\ \cite{Cucchieri}. Of course, other possibilities exist. One would be to average over the part of the residual gauge orbit which lies in the first Gribov region, yielding a Landau-Feynman-like gauge. Another would be to average over the complete residual gauge orbit. The latter is, however, highly non-trivial, but it should be possible to construct \cite{Hirschfeld:yq}. Further ones are discussed in sections \ref{sgf} and \ref{sdiscuss}. As the origin of field configuration space, and thus perturbation theory, is included in all these prescriptions, the correct reduction to perturbation theory, in the perturbative limit, is guaranteed.

Most interesting is whether and, if, to which extent results depend on this gauge choice. In particular, minimal Landau gauge is rather simple to implement in lattice simulations \cite{Cucchieri}, although some caveats will be discussed in the next section. On the other hand, absolute Landau gauge is an unambiguous definition of the gauge, and may therefore be implementable, in particular by means of condition \pref{propmin}, also in continuum calculations \cite{dse}. Hence a comparison of these two gauges is of great interest. Such comparisons have been performed repeatedly in the past in four dimensions \cite{sternbeck,Bogolubsky,Cucchieri:1997ns,mmp4,Silva:2004bv,Cucchieri:1997dx}. However, the results have been rather mixed, and no final result has been obtained so far, although effects seem likely, in particular in the case of the ghost propagator.

Therefore, fixing to the absolute Landau gauge is desirable. Unfortunately, fixing this gauge becomes exponentially harder with the number of lattice sites. Hence, a natural possibility to assess it is to investigate it in lower dimensions, where everything else applies as well, in particular the state of the results \cite{Zwanziger:2003cf,gzwanziger,Lerche:2002ep,Schleifenbaum:2004id,Huber:2007kc,Maas:2004se,Pawlowski:2003hq,Dudal:2008sp,Dudal:2008rm,Cucchieri:2007rg,limghost,Cucchieri:2008qm,Cucchieri:2003di,Cucchieri:2007ta,Maas:2007uv,Cucchieri:2006tf}. This will be done here, and it will turn out that this, in fact, is simpler and provides currently much more insight.

It has been argued that for expectation values of operators consisting of a product of a finite number of gluon field operators the results of the minimal and the absolute Landau gauge should coincide in the continuum and infinite volume limit \cite{Zwanziger:2003cf}\footnote{Because \pref{fumod} has a unique absolute minimum, by virtue of \pref{propmin} this can only occur in the sense of the difference becoming arbitrarily small, but an exact equality is never possible.}. This would apply to the gluon propagator, although not to the ghost propagator, which depends inversely on the gluon field. There is yet no indication that this is the case at reachable volumes, and also the results presented here do not support this conjecture. It has been argued that this may be a discretization artifact \cite{lvs}.

It should be noted that, due to the factor $p^{d-1}$ in the integral \pref{propmin}, the far infrared properties of the gluon propagator are likely irrelevant for the distinction between the minimal and the absolute Landau gauge, as, e.\ g., an effect at 100 MeV can be compensated by a $10^{d-1}$-times smaller effect at 1 GeV momentum: Any difference between the gluon propagators in the two gauges in the far infrared could be irrelevant to the distinction of the two gauges. This applies in particular to the value of the gluon propagator at zero momentum, which for a non-divergent gluon propagator is not contributing to the integral. In addition, due to asymptotic freedom the far ultraviolet is again irrelevant for the condition \pref{propmin}.

Furthermore, it should be noted that the statement \pref{propmin} is true for each configuration individually. The quantity on the right hand side of \pref{propmin} is the gluon propagator on each individual configuration. Nonetheless, the statement is also true on the average, but the distinction may be, and in fact is, only visible at very high statistics due to the $p^{d-1}$ factor and a difference in \pref{propmin} which is found to be of only about 0.3\% between the minimal and the absolute Landau gauge: For the results presented here, the maximization condition cannot be checked explicitly for the average gluon propagator due to the statistical noise. It is, of course, ensured for each configuration.

The method to fix the absolute Landau gauge is an evolutionary algorithm, which is detailed in appendix \ref{sabslan}. Note that similar algorithms have been used previously \cite{Oliveira:2003wa,Yamaguchi:1999hp}. The minimal Landau gauge will be fixed by an adaptive stochastic overrelaxation algorithm \cite{Cucchieri:2006tf,Cucchieri}.

\section{Gauge-fixing}\label{sgf}

There are a number of peculiarities when it comes to gauge fixing on the lattice, which have to be taken into account when dealing with Gribov copies. One of them is the fact that the complexity of gauge fixing is highly dependent on the configuration. A measure of the complexity is, e.\ g., the number of steps an iteration method needs to gauge fix a given configuration to the minimal Landau gauge. It turns out that this number varies by orders of magnitudes between different configurations. On the other hand, the extreme case of very complex configurations is rare. There is a relation to the so-called exceptional configurations \cite{sternbeck,Cucchieri:2006tf}, and hence the same name will be used. Furthermore, it turns out that the complexity also depends on the initialization of the iterative algorithm, i.\ e., the first guess for the correct gauge fixing. It is often possible to reduce the number of necessary sweeps considerably by choosing other initial conditions.

It is therefore tempting to restart the gauge-fixing procedure if the situation is exceptional, or even drop the corresponding configuration altogether, to improve computational efficiency. However, in general two different restarts correspond to two different Gribov copies, which is the observation on which many algorithms to fix to the absolute Landau gauge are based on \cite{Cucchieri:1997ns,Cucchieri:1997dx,sternbeck}. This may of course lead to a bias when sampling the residual gauge orbit, if the complexity of fixing the gauge is, e.\ g., related to the distance to the absolute minimum of \pref{fumod} along the gauge orbit. That such a bias possibly exists is indicated by the fact that a correlation between the value of the lowest eigenvalue of the Faddeev-Popov operator and the number of gauge fixing sweeps has been observed \cite{Cucchieri:2006tf}. The aim in this section is to either support this observation by searching for an impact on other observables as well, or to show that this is not the case. This will be done by performing measurements of observables as a function of the number of gauge-fixing sweeps necessary to fix to the minimal Landau gauge. Note that any of these statements apply only to the system investigated here, 40$^3$ at $\beta=4.2$, and may be different for different systems, although empirically these results seem to be typical.

To investigate this, configurations have been obtained from the Wilson action according to \cite{Cucchieri:2006tf} and then gauge-fixed to (minimal) Landau gauge according to \cite{Cucchieri:2006tf}. The details of the configurations are given in table \ref{tcgf} in appendix \ref{stech}. The gauge-fixing procedure was modified by imposing a limiting number of sweeps $n_s$. If during gauge-fixing more than this limit would be necessary, then the gauge-fixing process is restarted with another initialization. The latter was always a randomly chosen gauge transformation. If it was not possible to fix the gauge after twenty restarts, the configuration was dropped, and a new one was generated instead.

\begin{figure}
\includegraphics[width=\linewidth]{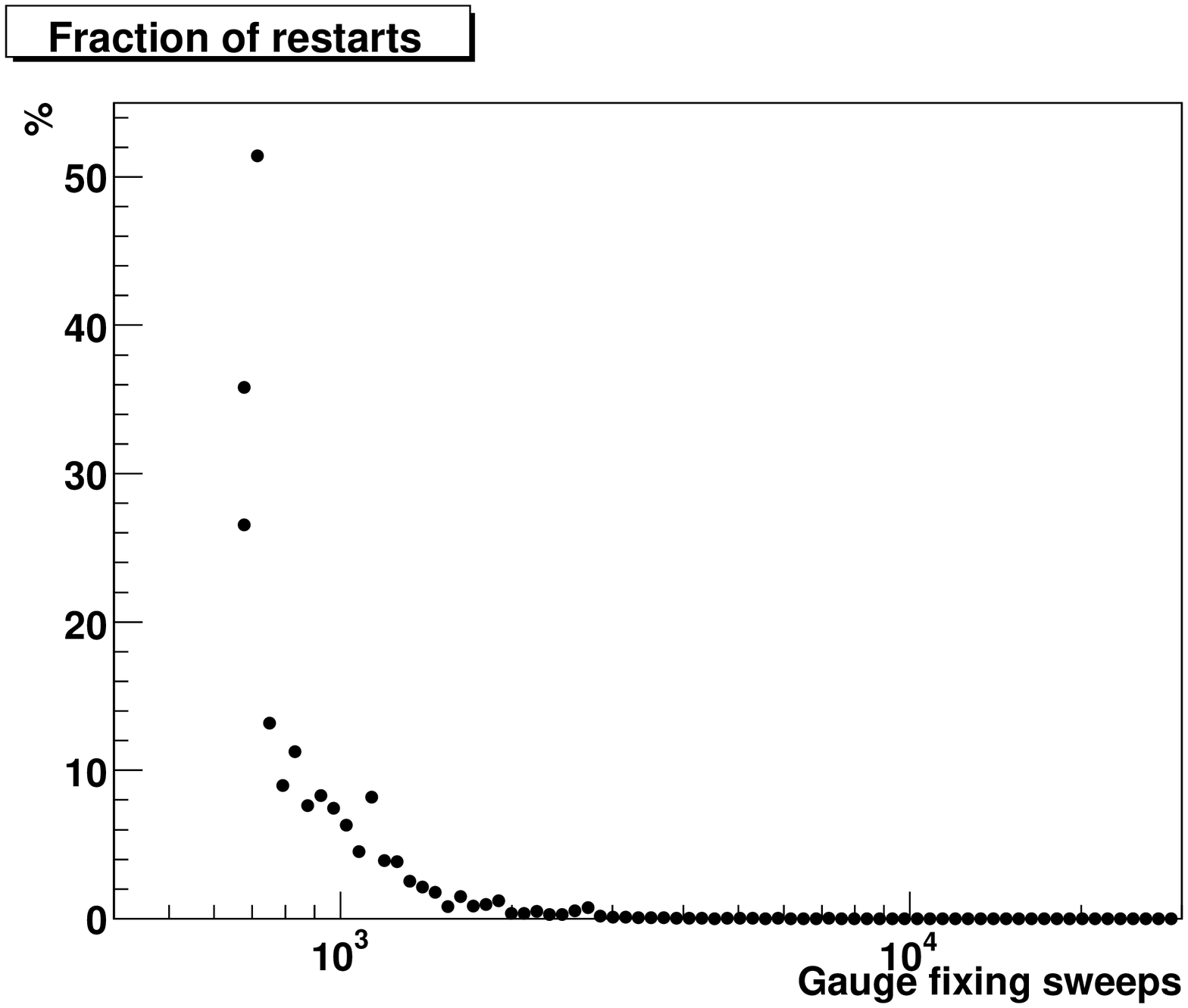}\\
\includegraphics[width=\linewidth]{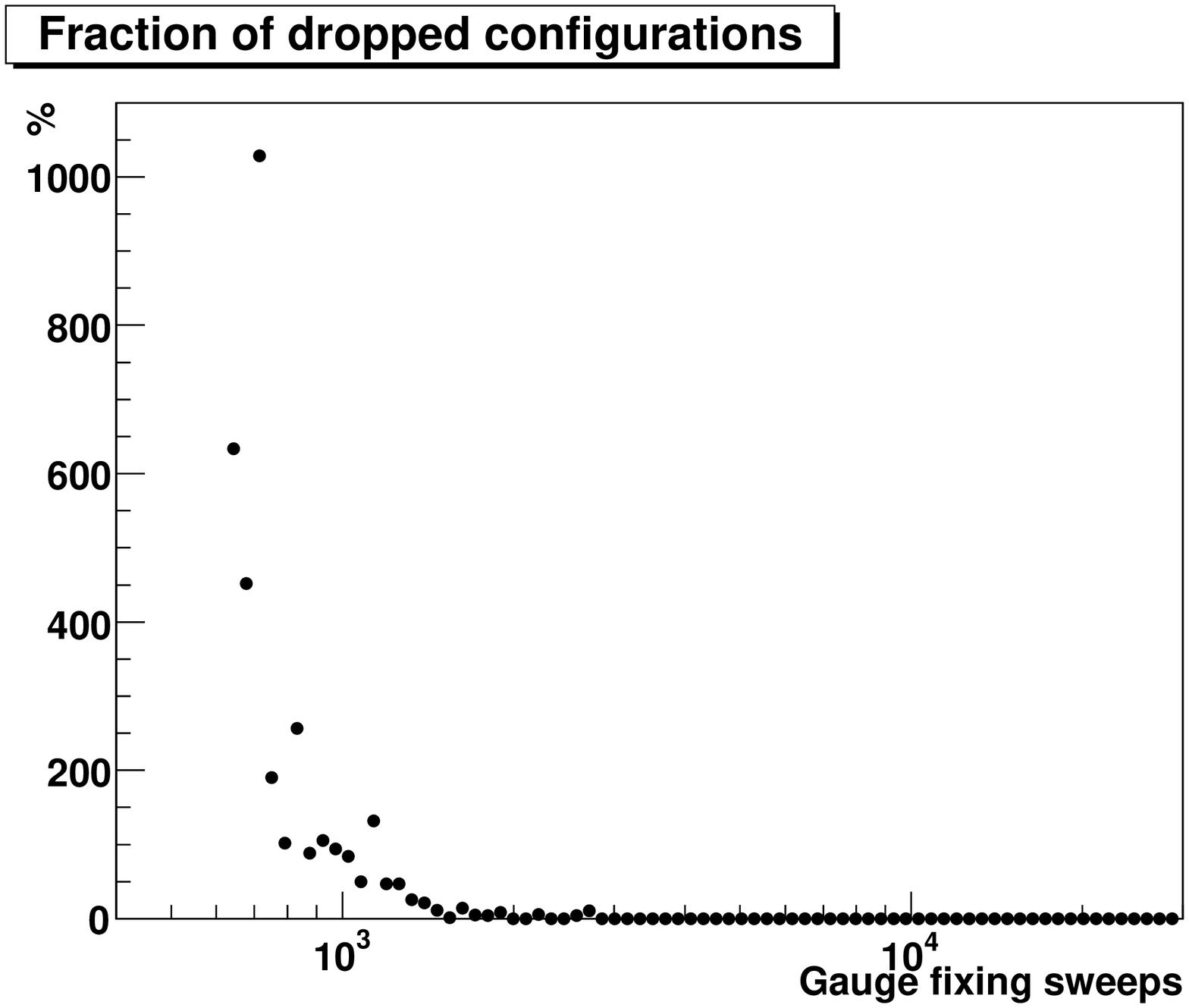}
\caption{\label{gf-rd}Top: Fraction of restarts of the maximum possible number of restarts during gauge-fixing as a function of $n_s$, not counting dropped configurations. Bottom: Fraction of dropped configurations compared to the total number of configurations as a function of $n_s$. Values greater than 100\% are possible, if more than every second configuration has been dropped.}
\end{figure}

The results have been compared to the case without any limitation, but otherwise the same gauge-fixing procedure. In this case, the maximum number of gauge-fixing sweeps observed was $n_s=28740$, while the average number was 1528(55), and the minimum number\footnote{This number was due to an optimization of the gauge-fixing procedure, and is actually only an upper limit for the minimum number.} 401. To quantify the amount of exceptional configurations with respect to the limit $n_s$ the number of necessary restarts and of dropped configurations\footnote{Indications have been found that it is very likely that if one configuration is dropped the next one(s) will be dropped also. This would indicate that the correlation time increases for configurations which are harder to gauge-fix, as the number of sweeps between two gauge-fixed measurements is sufficient for many other observables to decorrelate. To make this a more reliable statement more investigation is needed, however \cite{gcv}. To compensate this effect, multiple independent runs have always been performed throughout all results in this paper.} are shown in figure \ref{gf-rd}. It is visible that very extreme configurations are very rare. Within an order of magnitude of the maximum of gauge-fixing sweeps, it was always possible to find another Gribov copy which could be fixed sufficiently fast. Therefore, severe restrictions on the number of gauge-fixing sweeps are necessary for configurations to be dropped. Still, starting from a certain limit on, a very strong rise in both the number of restarts and the number of dropped configurations is observed.

It turns out that the exceptional configurations have no statistically significant impact on observables as long as the number of restarts is sufficiently small. As an example of the impact on measured quantities, the gluon propagator
\be
D_\mn^{ab}(x-y)=<A_\mu^a(x)A_\nu^b(x)>\nn
\ee 
\no and the ghost propagator
\be
D_G^{ab}(x-y)=-<(\pd_\mu(\pd_\mu\delta^{ab}-gf^{abc}A_\mu^c(x)))^{-1}>\nn
\ee 
\no will be used. In both cases, the color (and, for the gluon propagator, Lorentz) traced\footnote{None of the modifications of the gauge-fixing process performed here or in section \ref{sgauge} changes the fact that both propagators are color-diagonal \cite{Cucchieri:2006tf,Maas:2007uv}.} and Fourier-transformed propagators will always be shown, $D(p)$ and $D_G(p)$, respectively\footnote{A more precise definition can be found in \cite{Cucchieri:2006tf}, along with details on the determination of these quantities, including error determination.}.

\begin{figure}
\includegraphics[width=\linewidth]{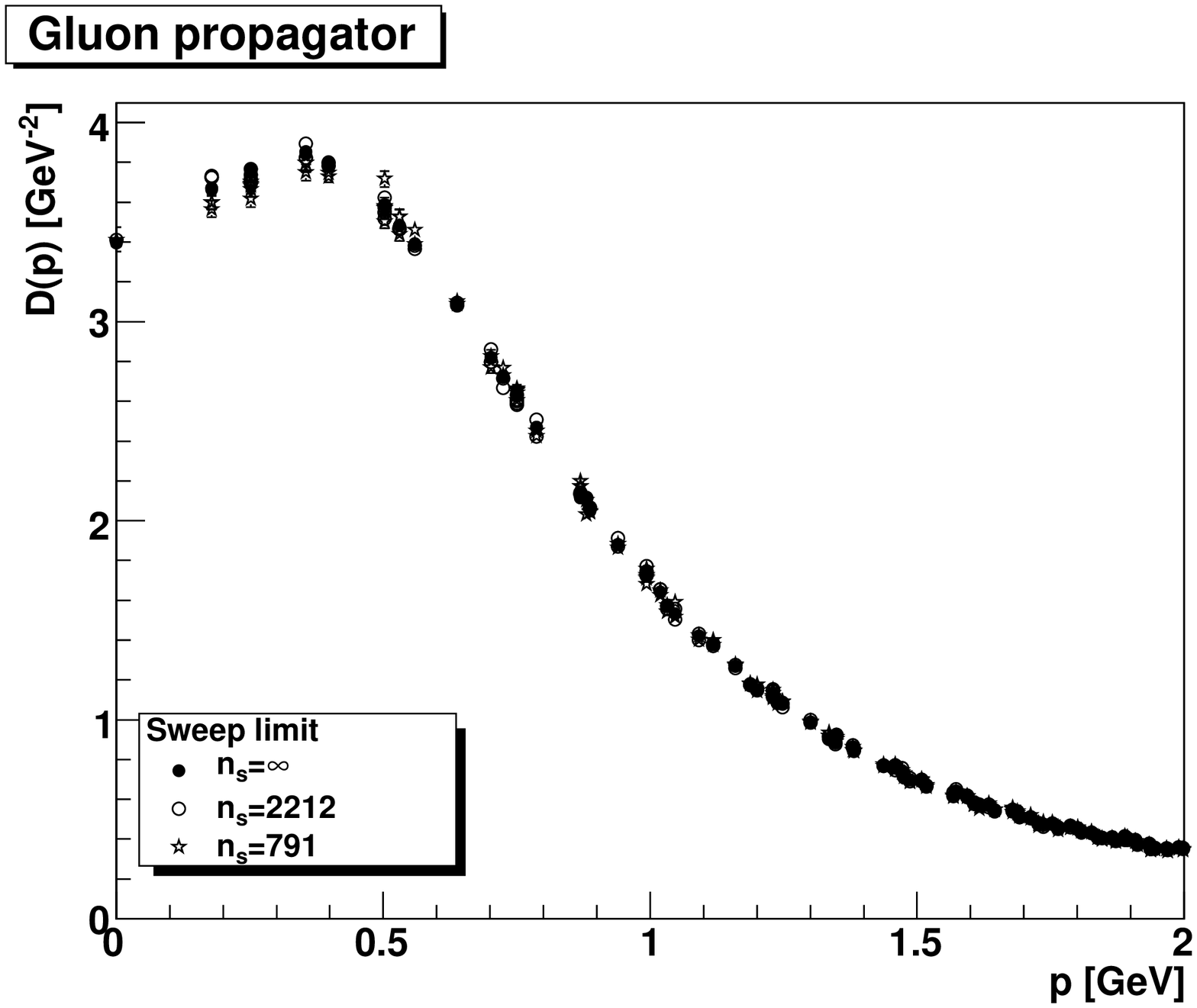}\\
\includegraphics[width=\linewidth]{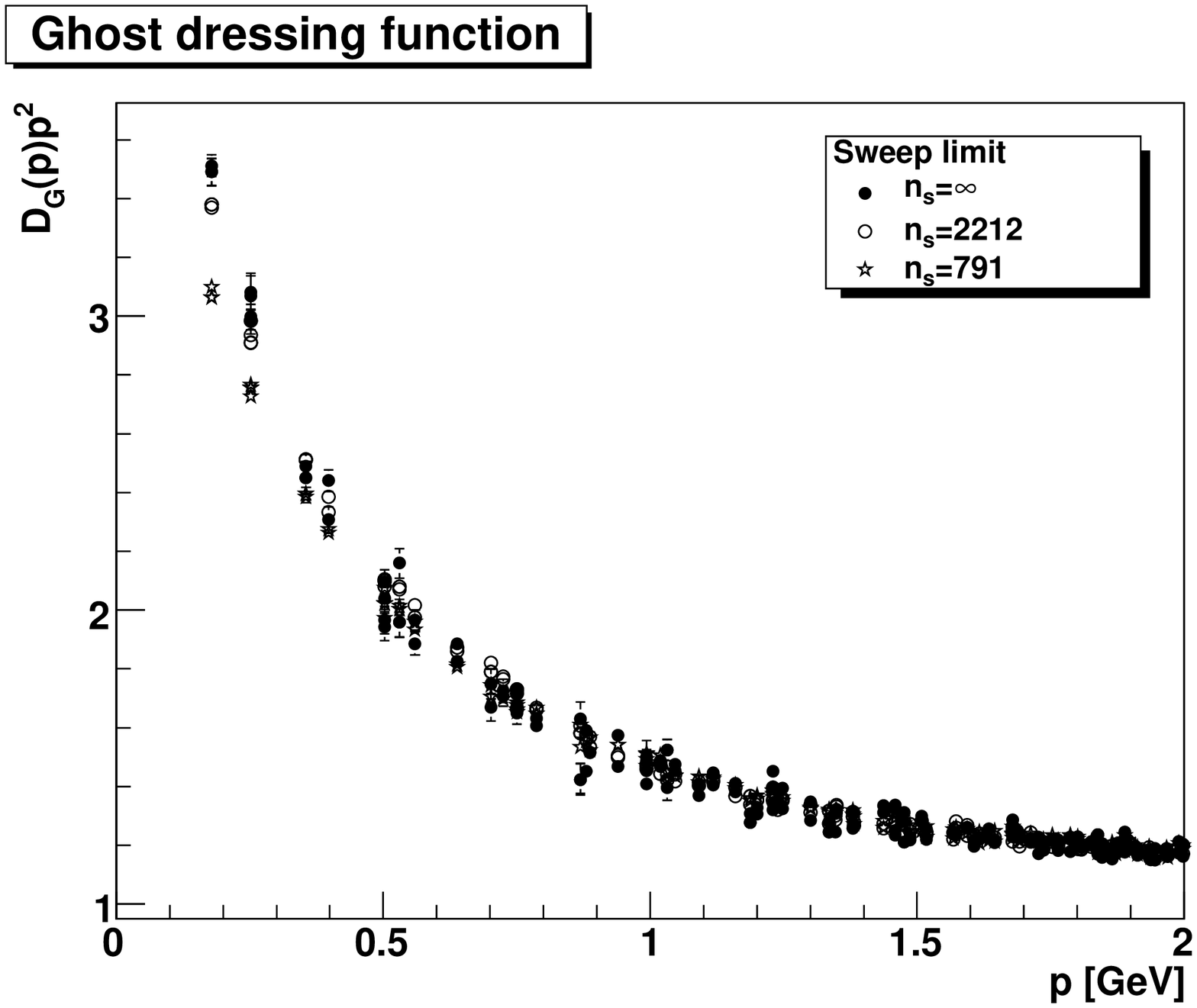}
\caption{\label{gf-prop}The gluon propagator (top panel) and the ghost dressing function (bottom panel) as a function of $n_s$ for $n_s=\infty$ (full circles), $n_s=2212$ (open circles) and $n_s=791$ (stars).}
\end{figure}

\begin{figure}
\includegraphics[width=\linewidth]{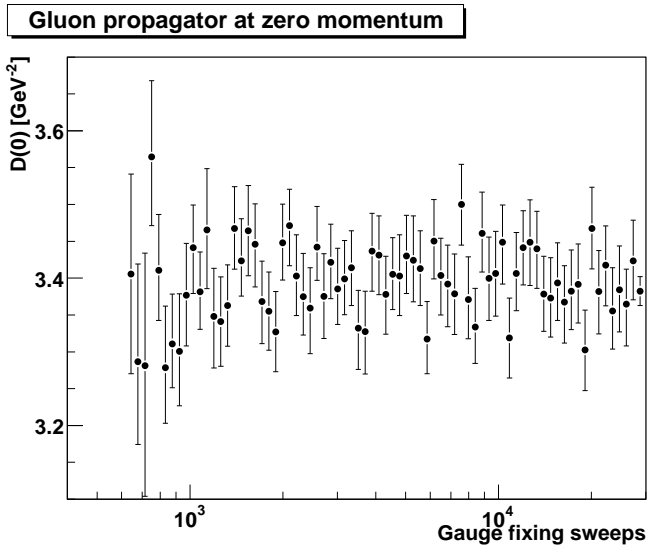}
\caption{\label{gf-d0}The gluon propagator at zero momentum as a function of the gauge fixing sweeps limit $n_s$.}
\end{figure}

The propagators are shown in figure \ref{gf-prop}. It is visible that there is essentially no change for the gluon propagator. In case of the ghost propagator the few lowest momentum points are affected, which would be in accordance with the observation of an influence on the lowest eigenvalue of the Faddeev-Popov operator \cite{Cucchieri:2006tf}. This latter influence is also observed here, manifesting itself by a systematic increase of the lowest eigenvalue of the Faddeev-Popov operator by about 30\%\footnote{The presented deformation of the sampling procedure of the residual gauge orbit would therefore influence the limits on the infinite volume ghost propagator constructed in \cite{limghost}, if it would be determined with the limits presented here.}. However, the lowest two momentum points of the ghost propagator are subject to finite volume effects \cite{Cucchieri:2007ta,Fischer:2007pf}, and the consequences are thus less relevant. Thus, at the respective level of biasing the selection of Gribov copies has almost no consequence within statistical errors for the propagators. As a further observable the gluon propagator at zero momentum will be investigated. This value is particularly important, as will be discussed in the next section. Its dependence on $n_s$ is shown in figure \ref{gf-d0}. Again, within statistical errors, there is no consequence of limiting the number of gauge-fixing sweeps.

Therefore, at least for a modest restart policy and at the given level of statistical errors, for the system investigated here, the consequences of the bias observed in \cite{Cucchieri:2006tf} have no relevant impact, except for the properties of finite volume effects. However, as will be seen in the next section, a bias in selecting Gribov copies can have consequences. Therefore, when desiring to use a limitation to improve the usage of computational resources, it would be necessary to make sure in advance whether for a given system and a given observable this introduces a bias. Hence, no such selection criteria will be used hereafter.

The reason for this effect may be that Gribov copies which are more simple to fix to the Landau gauge are in some sense more 'perturbative', which is by definition almost in Landau gauge. This would also explain the increase of the value of the lowest eigenvalue of the Faddeev-Popov operator. This is also supported by another observation: When the thermalization process is initialized with a cold, i.\ e.\ vacuum configuration, and not enough thermalization sweeps are performed to fully thermalize the configuration, it is usually much simpler to gauge fix than a fully thermalized configuration.

However, as there is an effect on correlation functions by performing this modification of the minimal Landau gauge it implies that 'minimal Landau gauge' is not by itself a sufficiently precise description of the gauge. Implicitly assumed is that dropping configurations is only a convenient numerical simplification, and is not altering the sampling of configurations. This is supported by the fact that the alterations observed have a smooth onset even when none or only a statistically insignificant number of configurations are dropped. A more precise description is hence needed to qualify what is meant by 'choosing randomly one representative inside the first Gribov region'. This has to be understood further to give a precise meaning to the concept of the minimal Landau gauge. Here, as an operational definition of the gauge the description of the gauge-fixing algorithm (with no limit on the number of sweeps) \cite{Cucchieri:2006tf} will be used.

This point will be discussed further in section \ref{sdiscuss}.

\section{Minimal vs. Absolute Landau gauge}\label{sgauge}

In the following the propagators, as examples of gauge-dependent Green's functions, in minimal and absolute Landau gauge will be compared. Aside from the general question to which extent gauge-dependent correlation functions actually differ in the two gauges, there is another reason to pose this question. Predictions exist from functional calculations on the infrared behavior of these correlation functions \cite{Zwanziger:2003cf,von Smekal:1997vx,Lerche:2002ep,Alkofer:2008jy}, in fact, for any correlation function \cite{Alkofer:2004it,Huber:2007kc}. These predict a power-law behavior of the correlation-functions in the far infrared, which is, if it exists, unique \cite{Fischer:2006vf}. In particular, the ghost propagator would be more singular than that of a massless particle, and the gluon propagator would be infrared vanishing if the ghost-gluon vertex were infrared constant \cite{Zwanziger:2003cf,von Smekal:1997vx,Watson:2001yv,Lerche:2002ep}. The latter is, inside this scenario, likely \cite{Schleifenbaum:2004id,Fischer:2006vf,Huber:2007kc}, and supported by lattice calculations \cite{Ilgenfritz:2006he,Cucchieri:2008qm,Maas:2007uv}. This applies to two, three and four dimensions. It has been discussed, whether this could be the form of the propagators in the absolute Landau gauge \cite{dse}.

However, as another possibility it has been suggested that the gluon propagator should be infrared finite and the ghost propagator behave like the one of a massless particle. This has been proposed in three and four dimensions \cite{Aguilar:2004sw,Binosi:2007pi,Aguilar:2007nf,Aguilar:2008xm,Boucaud:2006if,Boucaud:2008ji,Boucaud:2008ky,Dudal:2005na,Dudal:2007cw,Capri:2007ix,Dudal:2008sp,Dudal:2008rm}. In particular, this has been obtained in the context of a minimal Landau gauge \cite{Dudal:2008sp,Dudal:2008rm}.

The functional methods generally suffer from necessary assumptions and approximations to make them solvable. Although this necessity has been greatly reduced over time \cite{Alkofer:2004it,Huber:2007kc,Fischer:2006vf,dse}, the question which is the infrared behavior of the correlation functions still induces uncertainties. Therefore, it is natural to compare these results to lattice calculations, which have no need for assumptions. On the other hand, they suffer from artifacts due to finite volumes, discretization and numerical requirements. In particular, finite volumes have been a serious issue, especially in four dimensions \cite{Cucchieri:2007ta}. Recently, however, computational resources have become available, which reduced this problem considerable, at least for Yang-Mills theory with sufficiently coarse discretization \cite{cucchieril7,l72,Cucchieri:2007rg,limghost}.

A serious obstacle is still the necessity to fix to the absolute Landau gauge. This is a minimization problem in a high-dimensional space, therefore NP-complete, and thus exponentially hard. However, arguments have been made that at least for the gluon propagator in the limit of infinite volume in the continuum the results should be identical in the minimal and the absolute Landau gauge \cite{Zwanziger:2003cf}. Therefore, the numerically much simpler, since polynominally hard, minimal Landau gauge has been used to obtain results for the propagators on the lattice. These are in good agreement with an infrared finite gluon propagator and a massless ghost propagator in four and three dimensions \cite{cucchieril7,l72,Cucchieri:2007rg,limghost}, but an infrared vanishing gluon propagator and infrared enhanced ghost propagator in two dimensions \cite{Cucchieri:2007rg,limghost,Maas:2007uv}.

Still, the claim that both gauges coincide on the level of propagators has to be verified. Various attempts have been made, all on (necessarily) small volumes in four dimensions \cite{sternbeck,Bogolubsky,Cucchieri:1997ns,mmp4,Silva:2004bv,Cucchieri:1997dx}. Some discrepancies have been found for the ghost propagator between both gauges, which may or may not decrease with volume. However, as the problem increases with the dimensionality of the space, and thus lattice sites, it can be expected that it is much simpler in lower dimensions. Furthermore, it has been found that in two dimensions the results are in favor of a power-law solution \cite{Maas:2007uv}. It is therefore suggestive that also in higher dimensions in fact the power law solution would persist in the absolute Landau gauge. This is supported by the fact that in three dimensions, where the problem should be more severe than in two but less than in four, at least partly a power-law solution is observed \cite{Cucchieri:2003di}, and a discrepancy only appears at very large volumes \cite{Cucchieri:2007rg,limghost}. It is therefore a natural investigation to check the situation in two and three dimensions. This will be done here.

To this end, an evolutionary algorithm \cite{genetic} will be used, which is described in detail in appendix \ref{sabslan}. Still, this algorithm is not able to guarantee to fix to the absolute Landau gauge, as it cannot guarantee to find the absolute minimum of the gauge-fixing functional \pref{fumod}. This is a general problem \cite{sternbeck,Bogolubsky,Cucchieri:1997ns,mmp4,Silva:2004bv,Oliveira:2003wa,Yamaguchi:1999hp} for this type of minimization problem, which belongs to the same class as finding the ground state energy of spin-glass models. Only in the asymptotic limit of infinite computing time success can be guaranteed.

This has the unpleasant consequence that, even if the algorithm actually succeeded in finding the absolute minimum, it would be impossible to tell so. The only thing which can be hoped for is that the found gauge-fixed configuration is 'closer' to the absolute Landau gauge. Even a notion of 'closer' is impossible to define rigorously, as it is even unknown whether any type of gauge-fixing of the residual gauge orbit is smoothly connected to the absolute Landau gauge or not. That this is the case will be assumed henceforth.

Therefore, as a pragmatic ansatz, the fulfillment of the condition \pref{propmin}, finding the gluon propagator with the minimum total trace, will be taken as a criterion to compare the closeness to the absolute Landau gauge of two gauge-fixed configurations. Hence, the gauge-fixed configuration with the least value of ${\cal F}(A)$ found will be taken to be in the absolute Landau gauge. In addition, as a measure of the presence of Gribov copies the difference
\be
\Delta={\cal F}(A_\mathrm{worst})-{\cal F}(A_\mathrm{best})\label{delta}
\ee
\no will be used, where 'best' and 'worst' refer to the gauge-fixed field for a given configuration having the lowest and highest value of ${\cal F}(A)$, respectively.

It turns out that the function \pref{delta} has a generic shape, with potentially three, rather sharp, peaks. The lowest has values of $\Delta$ of the order of the numerical precision, and is therefore clearly due to numerical noise. The second peak clusters at very small values, typical of around the order of $10^{-15}$, likely due to lattice Gribov copies \cite{Giusti:2001xf}. The third peak is around $10^{-3}$ and appears to be related to continuum Gribov copies. With increasing volume, the population moves from the lowest to the highest peak. At sufficiently large volumes, only the peak at the highest value of $\Delta$ is populated. This could indicate that then at least one continuum Gribov copy is present in each configuration, in addition to any lattice copies.

Technical details on the lattice simulations performed can be found in appendix \ref{stech}.

\subsection{Two Dimensions}

The case of two dimensions is particularly interesting, as the results in lattice calculations and in continuum functional calculations agree \cite{gzwanziger,Maas:2007uv}. Furthermore, in this case the finite-volume effects for the gluon propagator behave in exactly the same way as in functional calculations, while they are apparently different for the ghost propagator \cite{Maas:2007uv,Fischer:2007pf}. It is therefore highly interesting, whether a difference in terms of correlation functions between the minimal and absolute Landau gauge exists, and, in particular, whether this will harm the agreement between the continuum and the lattice results.

\begin{figure}
\includegraphics[width=\linewidth]{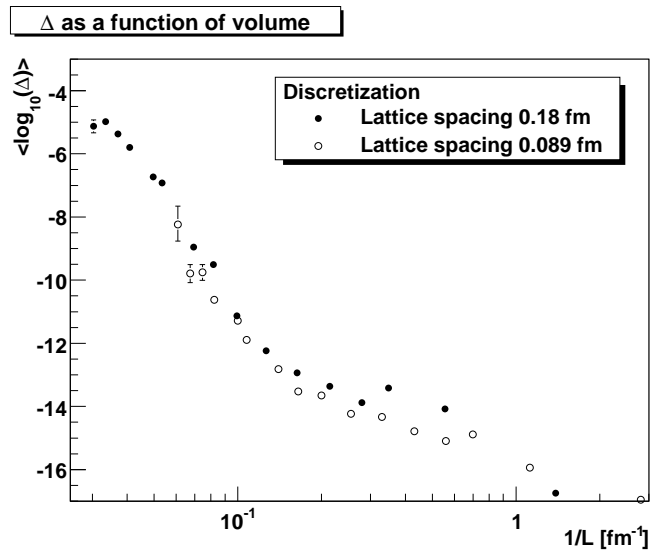}
\caption{\label{ld-2d}The presence of Gribov copies as measured by $\Delta$ as a function of inverse volume in two dimensions. Full circles correspond to $\beta=10$, while open circles correspond to $\beta=38.7$.}
\end{figure}

The dependence of $\Delta$ on the (inverse) volume is shown in figure \ref{ld-2d}. Even at quite large volumes, $\Delta$ is still rather small, implying that a considerable amount of the configurations do not yet turn up in the peak with the highest value of $\Delta$. This would indicate that in many cases the difference between the best and the worst Gribov copy is still negligible in terms of the gluon fields themselves. On the other hand, the exponential rise with volume is still clearly visible. It cannot, however, be anticipated at which level the function finally levels, and thus, whether Gribov copies could have a sizeable impact on the gluon fields at sufficiently large volumes. This remains to be seen. Furthermore, the discretization has only a small effect. In addition, this effect decreases with increasing volume.

\begin{figure}
\includegraphics[width=\linewidth]{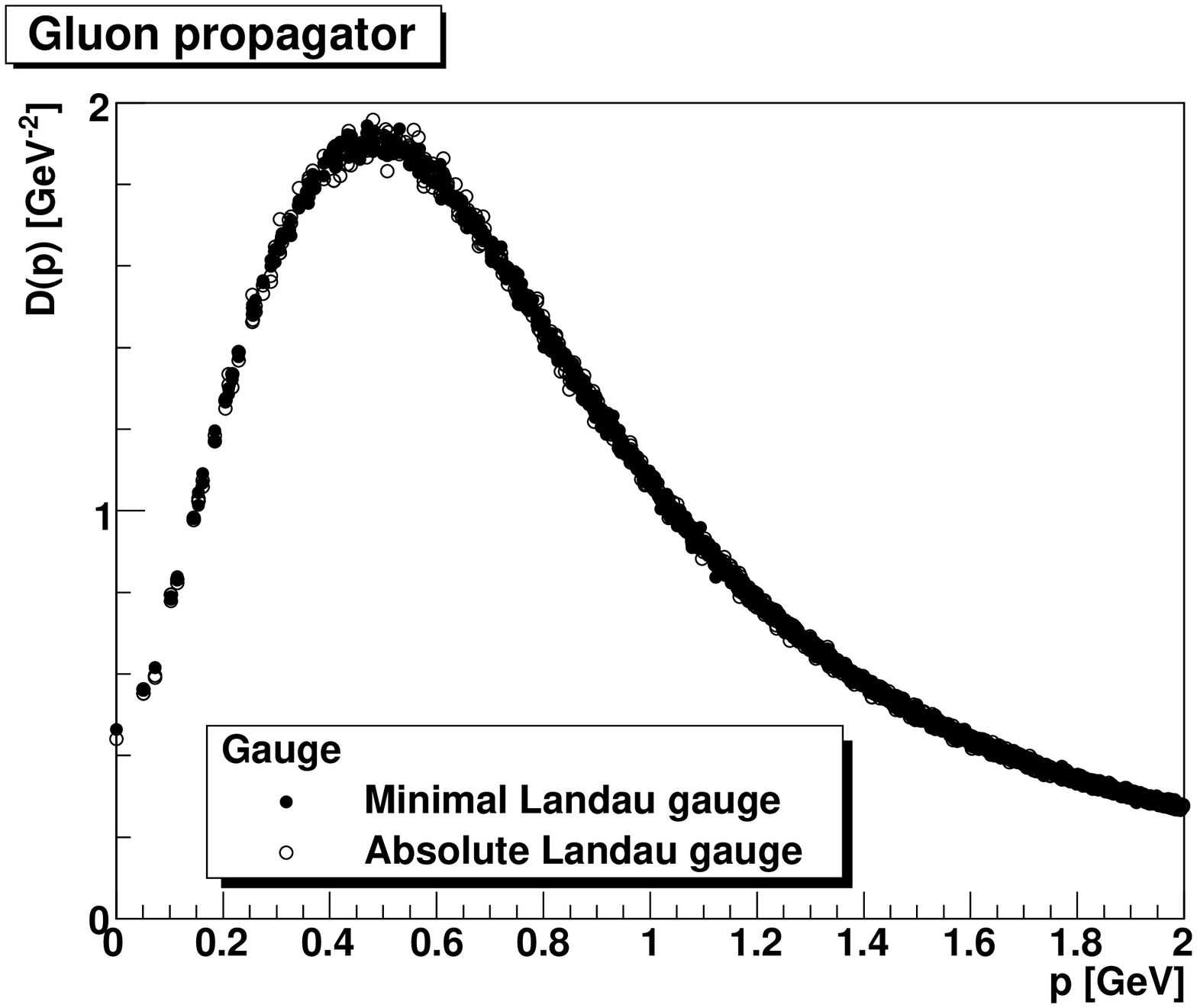}\\
\includegraphics[width=\linewidth]{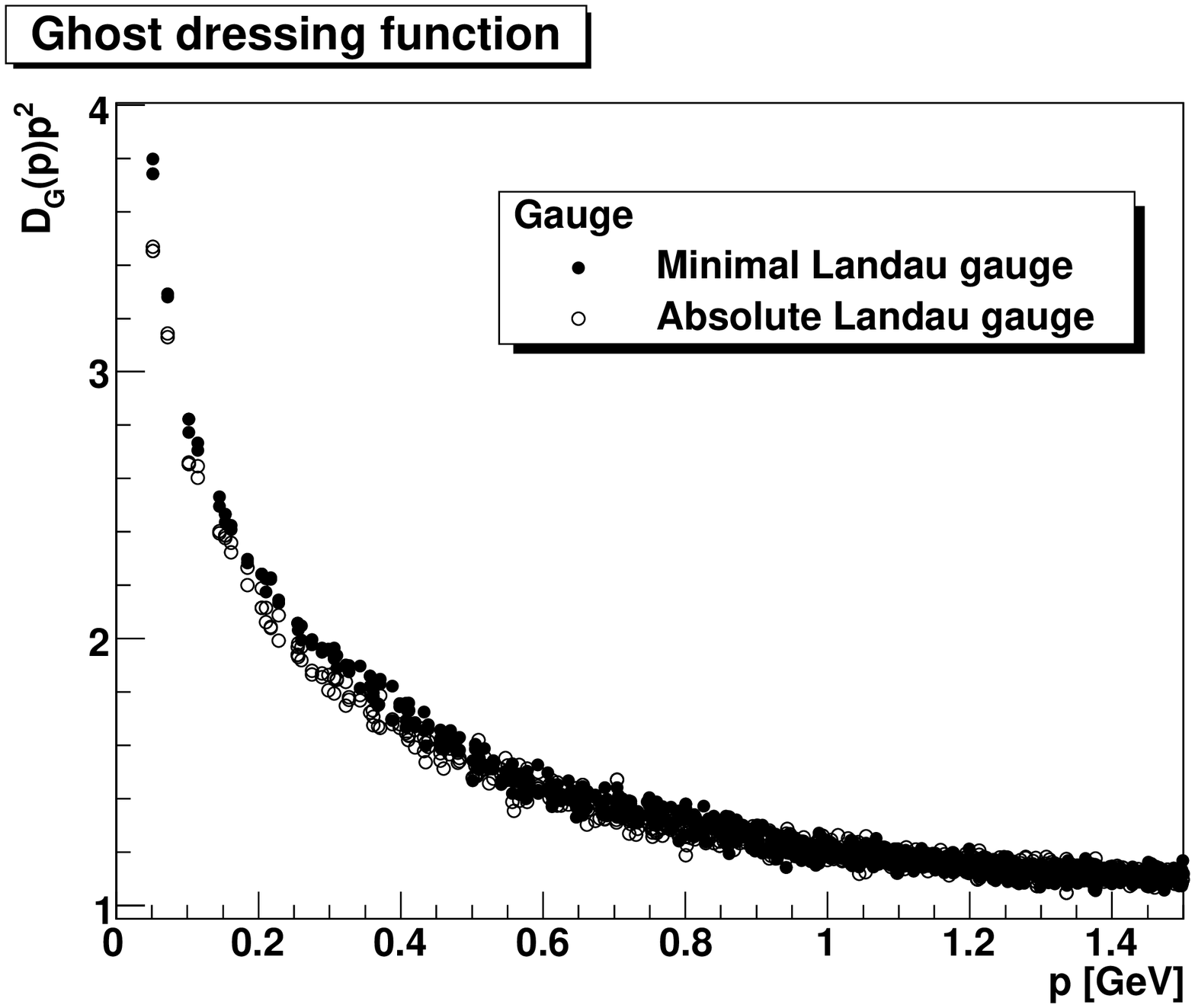}
\caption{\label{prop-2d}The propagators from a $136^2$ ((25 fm)$^2$) lattice at $\beta=10$. The top panel shows the gluon propagator and the bottom panel the ghost dressing function. Full circles correspond to the results in the minimal Landau gauge and open circles to the absolute Landau gauge.}
\end{figure}

\begin{figure}
\includegraphics[width=\linewidth]{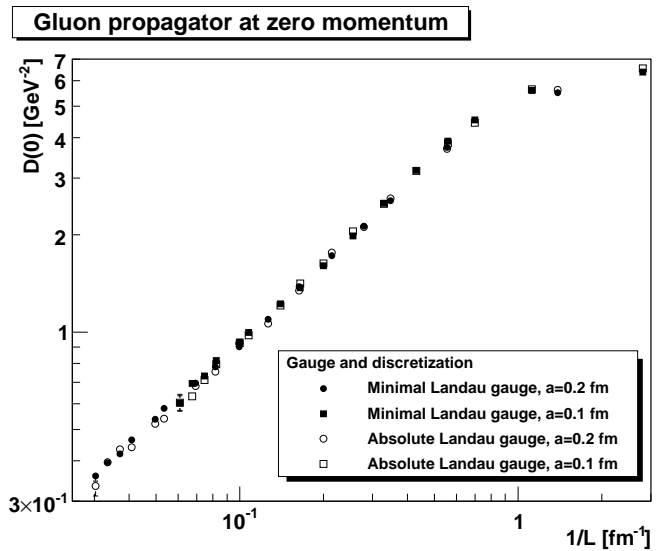}
\caption{\label{d0-2d}The gluon propagator at zero momentum as a function of inverse volume. Full and open symbols correspond to the minimal and absolute Landau gauge, respectively. Circles are from calculations at $\beta=10$ and squares from calculations at $\beta=38.7$.}
\end{figure}

However, a statement on the average difference between the field variables is not sufficient for a statement on their correlation functions. Therefore, these are studied separately. For a very large volume, the difference between the propagators in minimal Landau gauge and in absolute Landau gauge are shown in figure \ref{prop-2d}. The impact is quantitatively rather small for the propagators in the two different gauges. A statistically significant effect is only seen for the ghost propagator at momenta below 250 MeV. There, the propagator in the absolute Landau gauge is systematically smaller than in minimal Landau gauge. However, the effect is strongest for the two lowest momentum points, which are anyhow unreliable due to finite volume effects. There is also a small effect on the border of statistical significance for the gluon propagator. The gluon propagator in absolute Landau gauge is smaller in the infrared than the one in minimal Landau gauge. This would be in agreement with the condition \pref{propmin}. Nonetheless, the strongest effects are in the domain where finite volume artifacts dominate. Therefore, the effect is almost negligible. Still, the volume dependence of the gluon propagator at zero momentum is a quantity for which predictions exist \cite{Fischer:2007pf}, and which in the minimal Landau gauge are fulfilled \cite{Maas:2007uv}. In particular, it behaves as a power-law and vanishes in the infinite-volume limit. This is not altered in the absolute Landau gauge, as seen in figure \ref{d0-2d}. As can be seen, there is essentially no difference in the volume dependence.

\begin{figure}
\includegraphics[width=\linewidth]{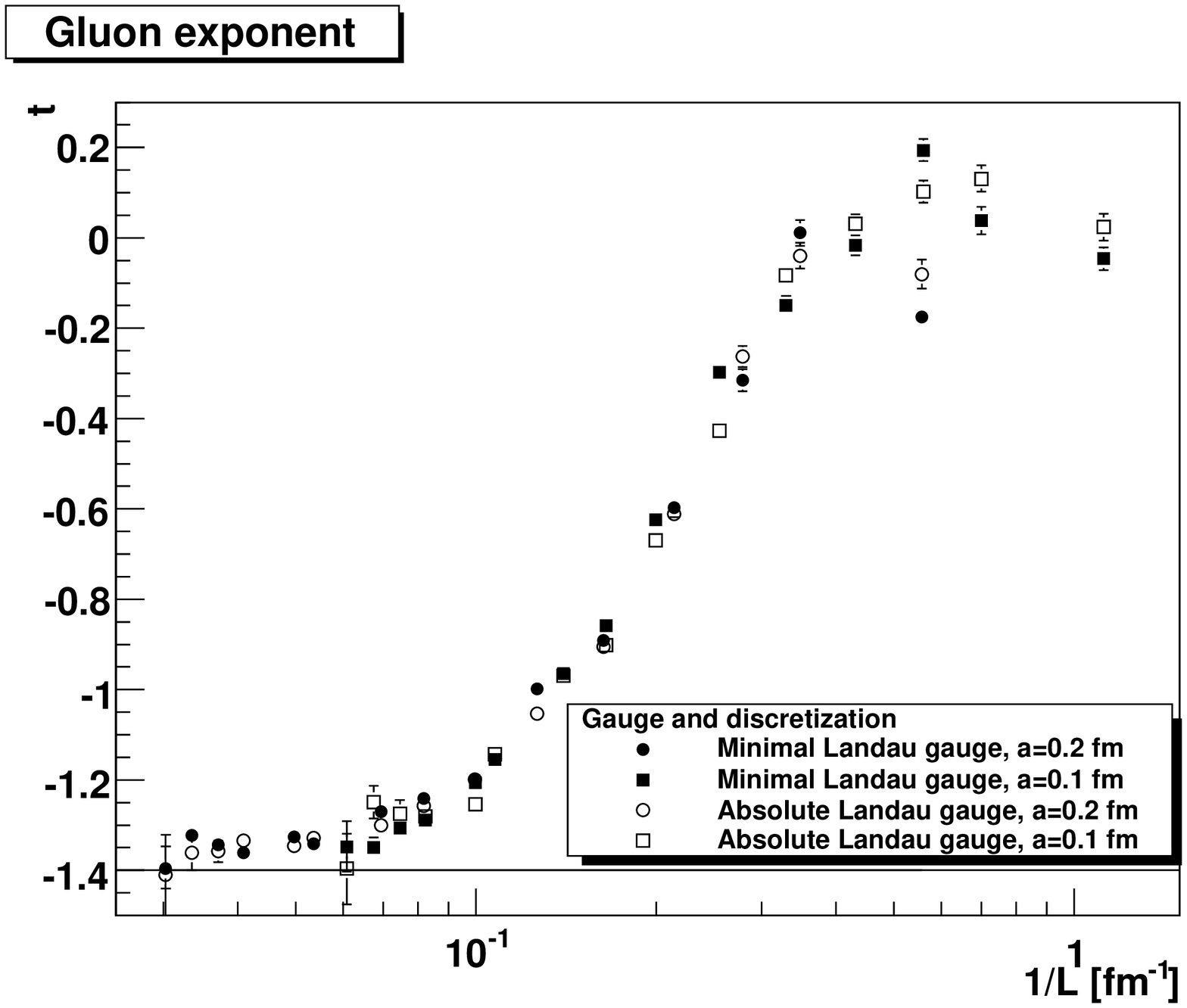}\\
\includegraphics[width=\linewidth]{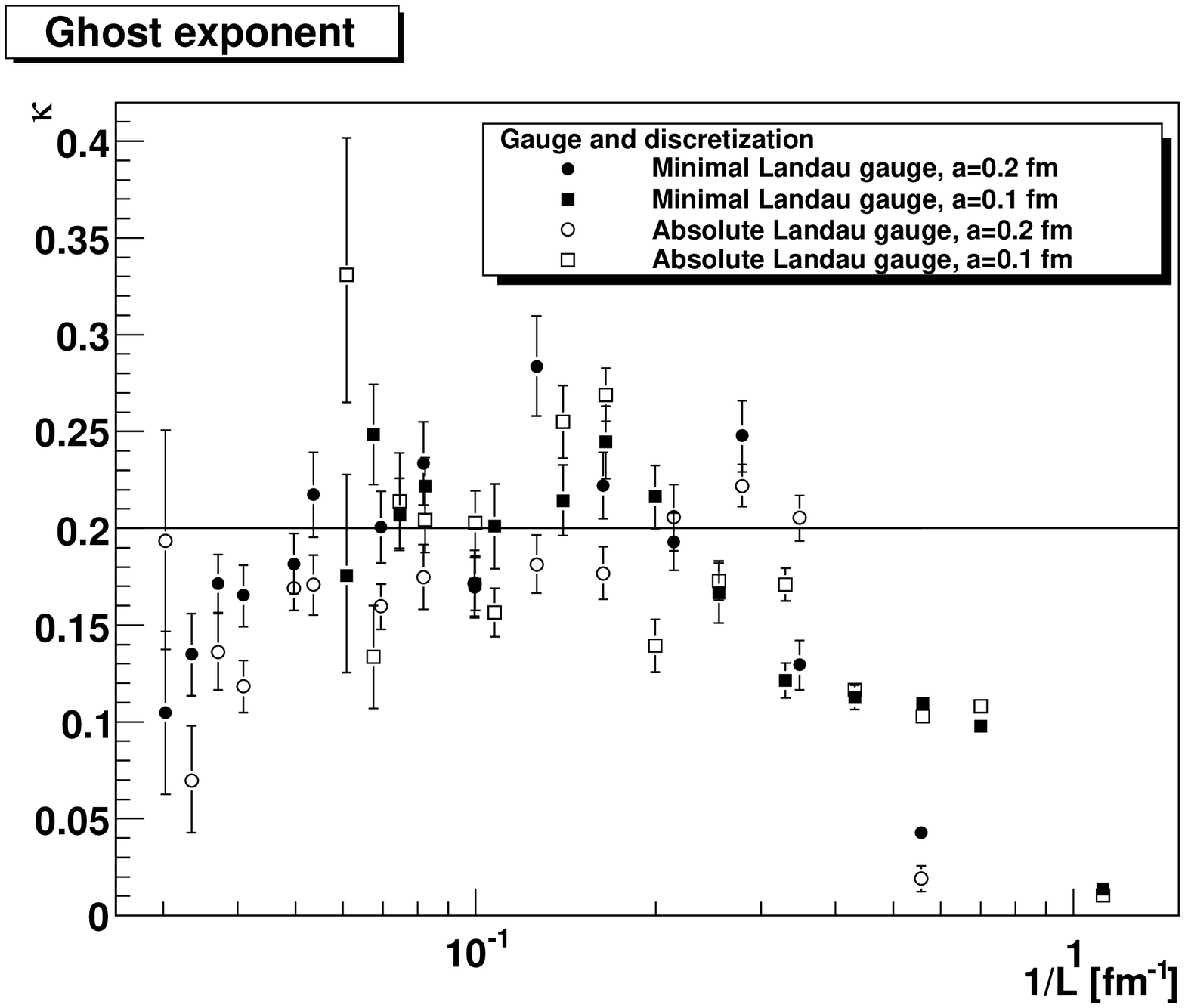}
\caption{\label{d0-exp}The gluon exponent $t$ (top panel) and the ghost exponent $\kappa$ (bottom panel) as a function of volume. Full and open symbols correspond to the minimal and absolute Landau gauge, respectively. Circles are from calculations at $\beta=10$ and squares from calculations at $\beta=38.7$. The lines indicate the infinite-volume predictions from functional calculations \cite{gzwanziger}.}
\end{figure}

A final possibility for comparison are differential characteristics of the propagators. Both propagators are predicted to behave as power-laws in the far infrared region \cite{gzwanziger,Lerche:2002ep}. This can be parameterized by
\bea
D(p)&\approx_{p\ll \Lambda_{\mathrm{YM}}}&A_Z(p^2)^{-1-t}\nn\\
D_G(p)&\approx_{p\ll \Lambda_{\mathrm{YM}}}&A_G(p^2)^{-1-\kappa},\nn
\eea
\no with constant prefactors $A_Z$ and $A_G$. In particular, the effective infrared exponents $t$ and $\kappa$ as a function of volume are interesting, as they are the key quantities in comparison to functional calculations. This volume dependence of the exponents, determined in the same manner as in \cite{Maas:2007uv}, is shown in figure \ref{d0-exp}. The gluon exponent $t$ does not differ in both gauges, and in both cases it tends towards the predicted value in the infinite volume limit. In case of the ghost exponent, there seems to be a systematic difference, both in central value and in volume dependence between both gauges. However, this fact is not statistically significant\footnote{With the statistics employed in \cite{Maas:2007uv}, this difference would not have been visible at all.}. In addition, in both gauges the exponent is compatible to the predicted value, within statistical accuracy. The variations of the central values for the two largest volumes indicate that the results at the respective level of statistics become already unreliable, a consequence of the fit procedure \cite{Maas:2007uv}.

\begin{figure}
\includegraphics[width=\linewidth]{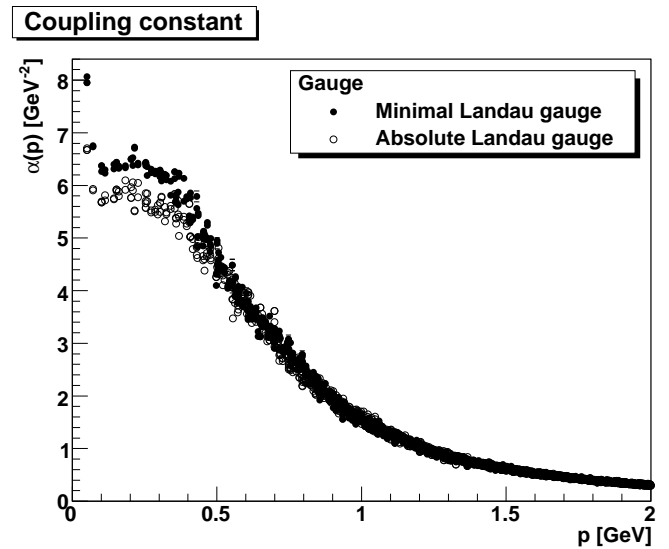}
\caption{\label{alpha-2d}The quantity \pref{alpha} from a $136^2$ ((25 fm)$^2$) lattice at $\beta=10$. Full circles correspond to the results in the minimal Landau gauge and open circles to the absolute Landau gauge. Note that the two lowest momentum points are affected by finite volume artifacts \cite{Maas:2007uv}.}
\end{figure}

A more sensitive test is the relation predicted between both exponents \cite{gzwanziger,von Smekal:1997vx,Lerche:2002ep,Pawlowski:2003hq}
\be
t=-2\kappa-\frac{4-d}{2}.\label{exprel}
\ee
\no As a consequence, the quantity 
\be
\alpha(p)=D(p)D_G(p)^2p^{2+d}\label{alpha}
\ee
\no should be a constant in the far infrared. This quantity is proportional to a possible definition of the coupling constant \cite{Lerche:2002ep} and can be regarded as an effective coupling constant.

The relation \pref{exprel} holds in both gauges, as can be seen from the quantity \pref{alpha} shown in figure \ref{alpha-2d}. However, the constant of proportionality shows a difference between both gauges. This implies that the difference observed for the ghost propagator in both gauges in figure \ref{prop-2d} is likely rather due to a change in the prefactor $A_G$ from gauge to gauge than in the value of the exponent $\kappa$.

Hence, it seems that there is at best a quantitative difference between the propagators in both gauges, but not a qualitative one. Even this quantitative difference seems to be rather small. In particular, the propagators in both gauges agree with the predictions from functional calculations in the continuum. However, after investigating the case in three dimensions in the next section, it will be necessary to reconsider this conclusion. This will be done in section \ref{ssint}.

\subsection{Three Dimensions}

In three dimensions, the situation is quite different than in two dimensions. The predictions from functional calculations are qualitatively identical \cite{gzwanziger,Lerche:2002ep,Maas:2004se,Pawlowski:2003hq}. Results on small and medium-sized lattices seem to confirm these predictions \cite{Cucchieri:2003di,Cucchieri:2006tf,Cucchieri:1999sz}. However, results on the largest lattices suggest rather an infrared finite gluon propagator and an infrared tree-level-like ghost propagator \cite{Cucchieri:2007rg,limghost}, as obtained in functional calculations in \cite{Dudal:2008rm}. Still, at least in an intermediate momentum window these results exhibit a power-law-like behavior. The latter observation is not a finite volume effect.

\begin{figure}
\includegraphics[width=\linewidth]{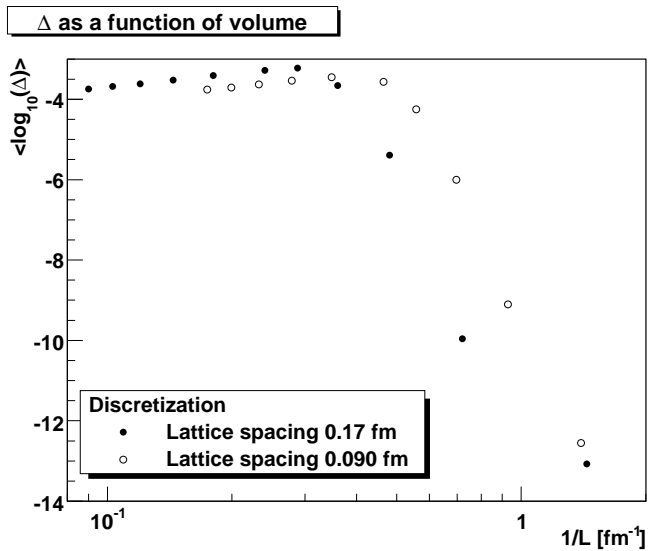}
\caption{\label{ld-3d}The presence of Gribov copies as measured by $\Delta$ as a function of inverse volume in three dimensions. Full circles correspond to $\beta=4.24$, while open circles correspond to $\beta=7.09$.}
\end{figure}

Starting again with the measure for the presence of Gribov copies $\Delta$, the result is already drastically different from the two-dimensional case. This is shown in figure \ref{ld-3d}. As in the two-dimensional case, an exponential rise with volume is visible. However, it saturates at a value of about $10^{-3}$, corresponding to the highest-valued peak in the distribution of $\Delta$. In fact, at the largest volumes, $\Delta$ is confined to a rather small region around this peak, implying that in all configurations there are significant differences in terms of ${\cal F}(A)$ between the best and worst Gribov copy. If these large differences in fact correspond to continuum Gribov copies, this implies that, starting from volumes of the order of about (3 fm)$^3$, Gribov copies play an important role.

\begin{figure}
\includegraphics[width=\linewidth]{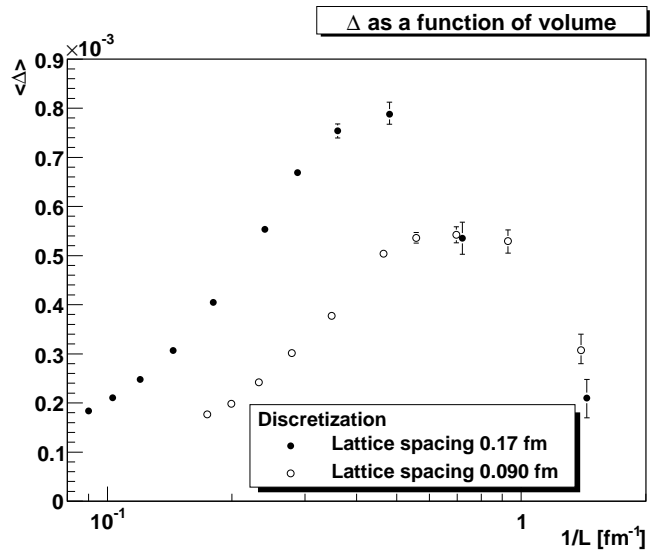}
\caption{\label{d-3d}The presence of Gribov copies as measured by $\Delta$ as a function of inverse volume in three dimensions, averaged on a linear scale. Full circles correspond to $\beta=4.24$, while open circles correspond to $\beta=7.09$.}
\end{figure}

A little detail is that $\Delta$ seems to decrease once more when the volume becomes larger. This is more drastically seen when averaging $\Delta$ on a linear scale, as shown in figure \ref{d-3d}. Here, it is visible that $\Delta$ starts to decay significantly with increasing volume. On the one hand, this can indicate that the local minima become more and more degenerate with the absolute minimum, and is thus a statement on the nature of Gribov copies\footnote{This happens, e.\ g., in 2d U(1) gauge theory \cite{deForcrand:1994mz}. I am grateful to L.~von Smekal for pointing out this possibility.}. It could also imply that the gauge-fixing algorithm becomes unable to find a good approximation of the absolute minimum. This would be possible if the search space covered is not sufficiently enlarged to compensate for the exponential increase of the number of Gribov copies with volume. This possibility is discussed in more detail in appendix \ref{sabslan}, and is the most likely one. It could, of course, also be a mixture of both effects. Anyway, it may potentially be affecting the reliability of the results at sufficiently large volumes. Eventually, if the computational resources are not increased exponentially with volume as well, the gauge-fixing algorithm will fail. As a consequence, the result will become indistinguishable from the ones in minimal Landau gauge. This may be the onset of it. A third possibility is that the Zwanziger conjecture may be correct in the end: A decrease in $\Delta$ implies that the integrated strength of the gluon propagator becomes more and more similar between minimal and absolute Landau gauge once more, leading possibly eventually to coinciding gluon propagators. However, it will be shown below that this is not equivalent to a point-wise coincidence of the gluon propagators yet, although some degradation is observed for the results in the largest volume.

\begin{figure*}
\includegraphics[width=0.5\linewidth]{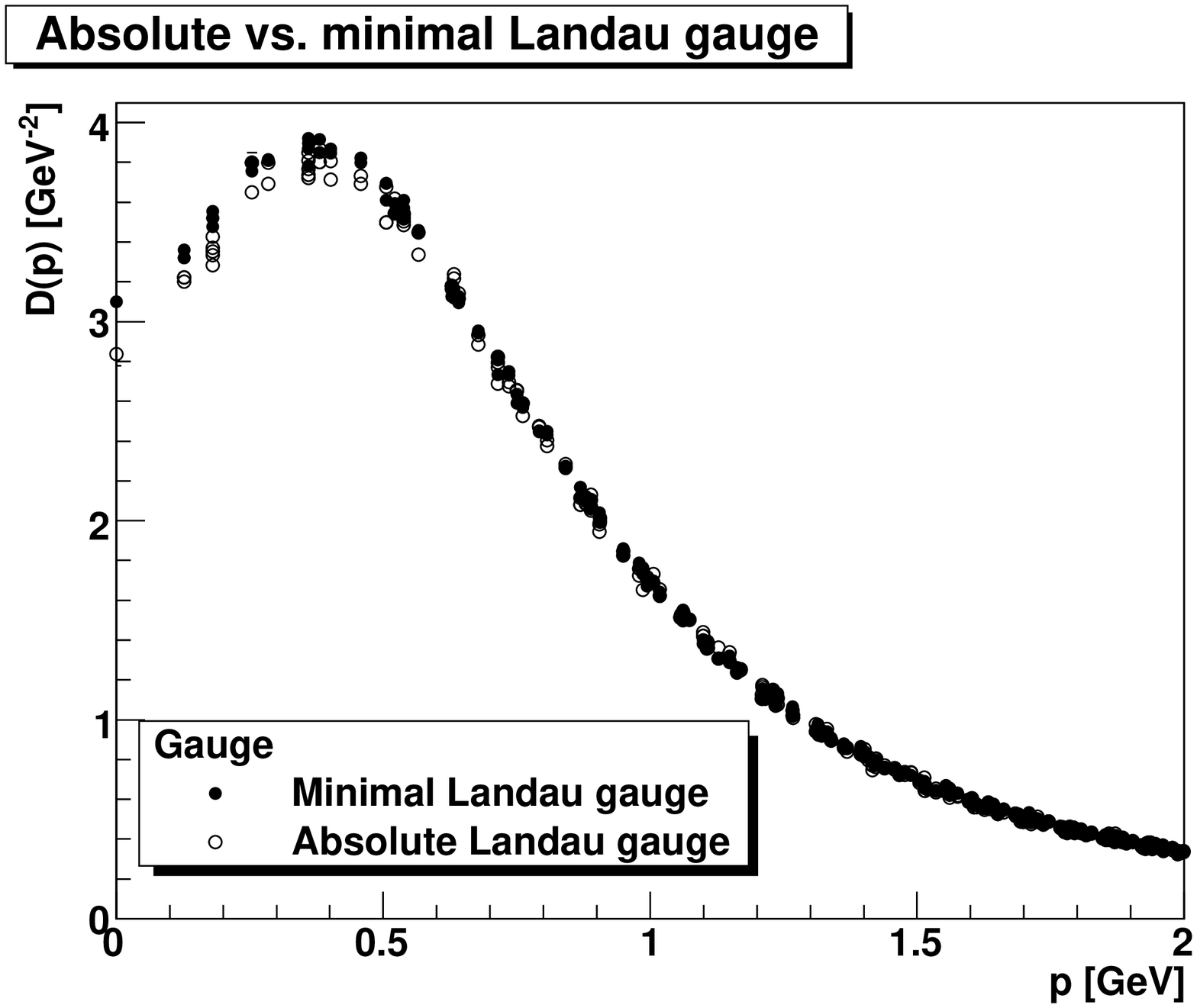}\includegraphics[width=0.5\linewidth]{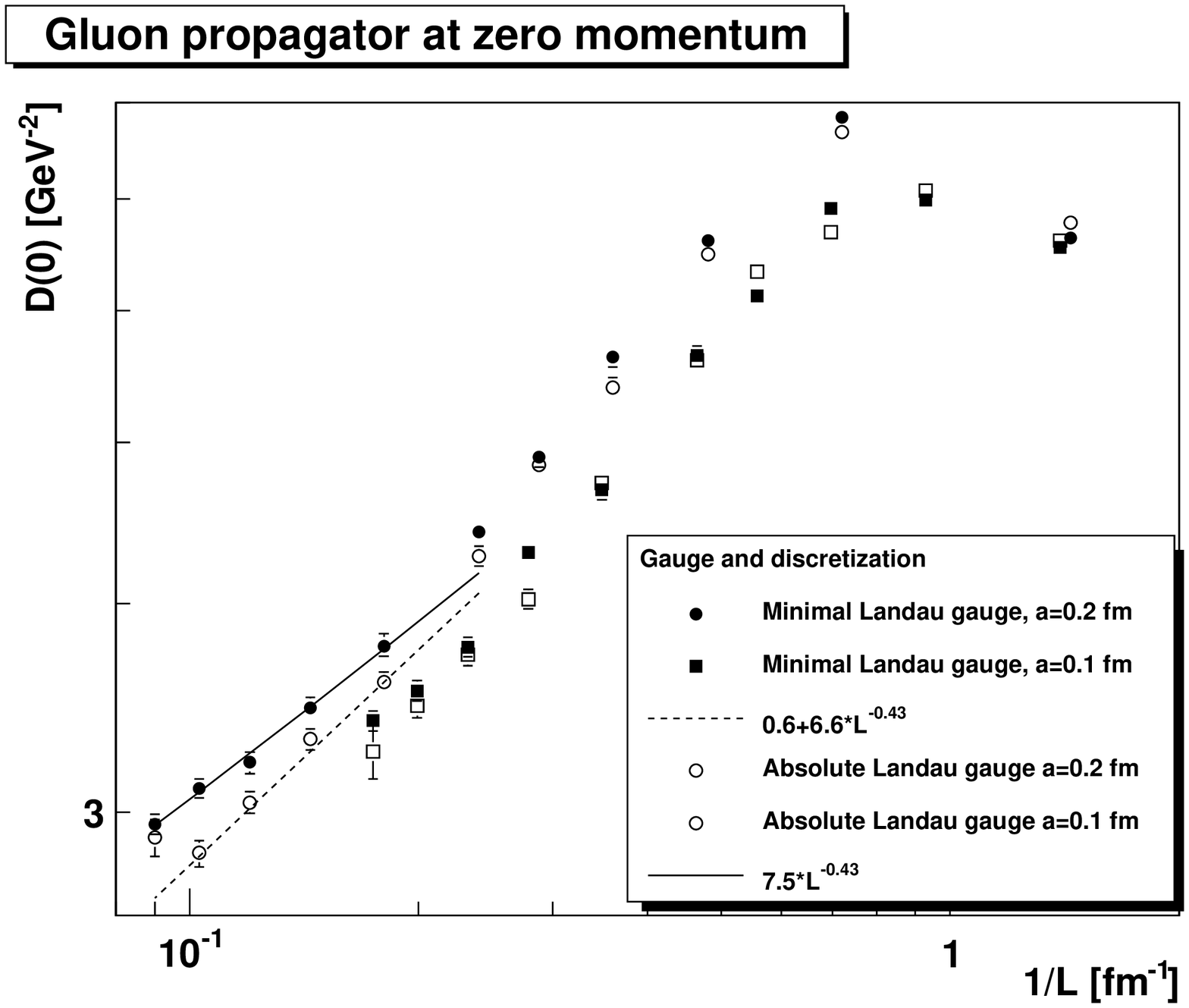}\\
\includegraphics[width=0.5\linewidth]{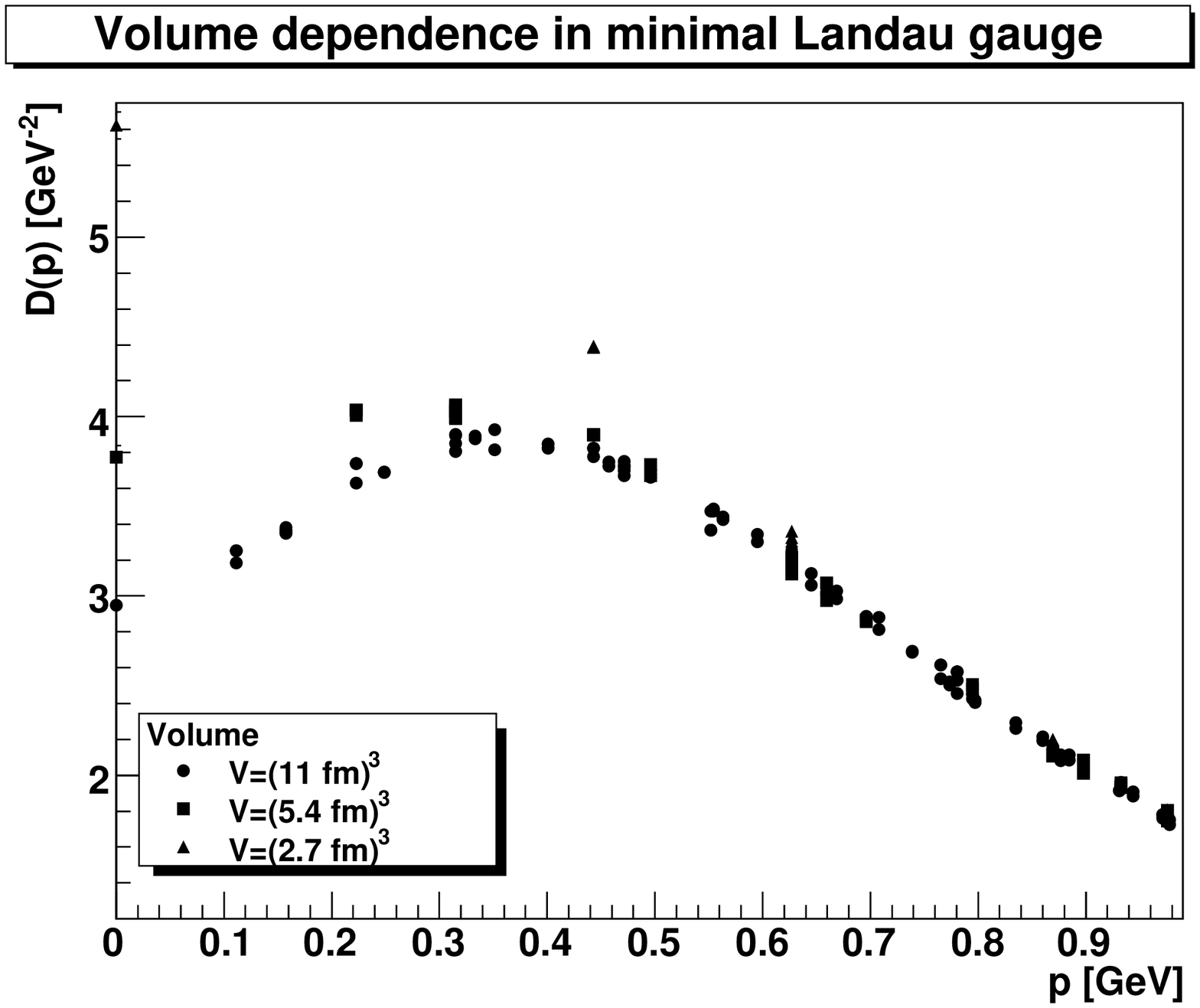}\includegraphics[width=0.5\linewidth]{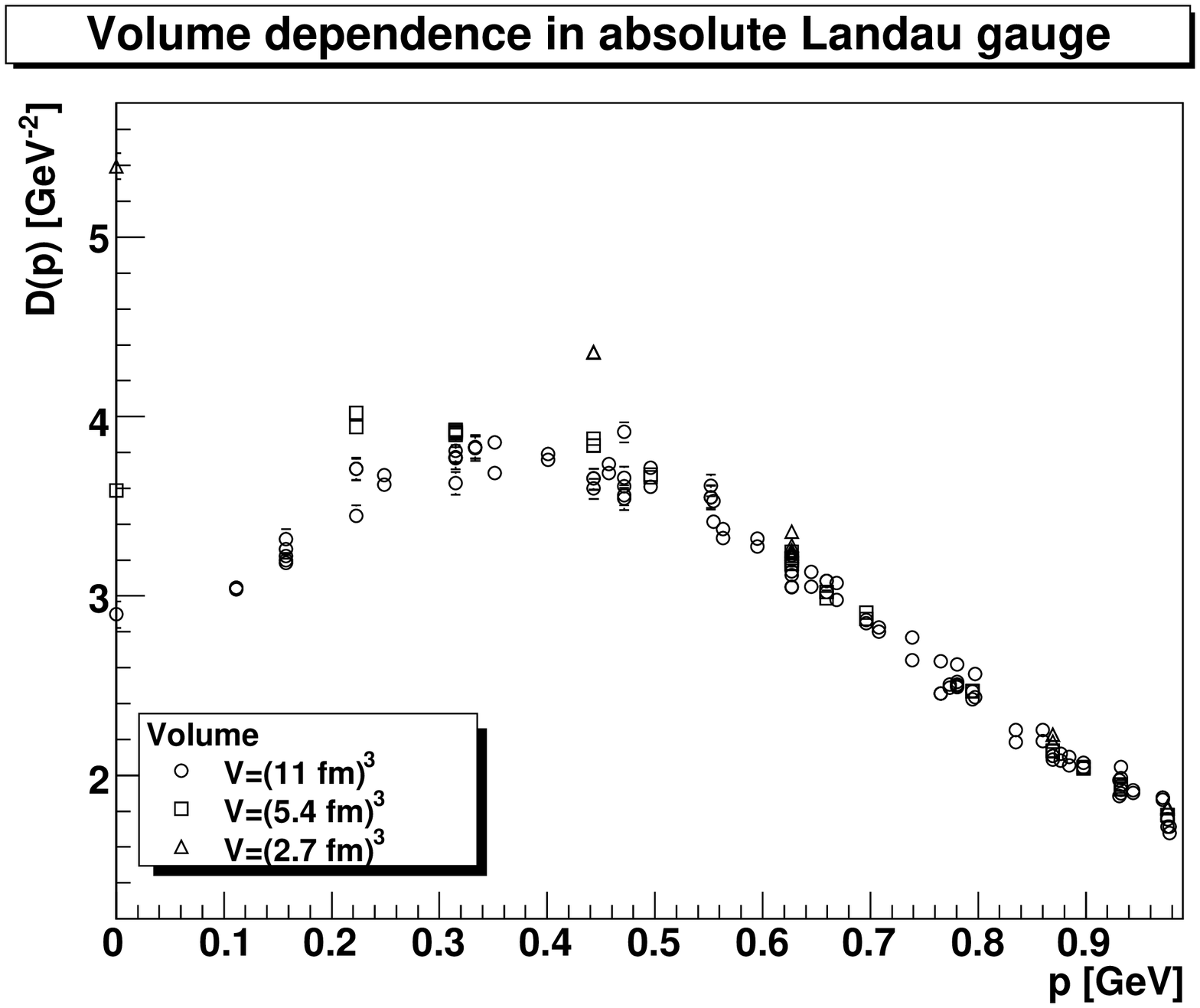}
\caption{\label{gp-3d}The gluon propagator in three dimensions. The top left panel shows the comparison between minimal (full circles) and absolute (open circles) Landau gauge on a $56^3$ lattice at $\beta=4.24$, corresponding to $V=(9.5$ fm$)^3$. The bottom panels show the volume dependence for the gluon propagator in minimal (left panel) and absolute (right panel) Landau gauge. Circles, squares, and triangles correspond to $V=(11$ fm$)^3$, (5.4 fm)$^3$, and (2.7 fm)$^3$, respectively, all calculated at $\beta=4.24$. The top right panel shows the gluon propagator at zero momentum, together with fits for the coarser discretization. Full symbols are from the minimal Landau gauge, while open symbols are from the absolute Landau gauge. Circles and squares correspond to calculations at $\beta=4.24$ and $\beta=7.09$.}
\end{figure*}

Monitoring the gluon propagator is another possibility to measure the success of the gauge-fixing procedure. The result for the gluon propagator in both gauges is shown in one large volume in figure \ref{gp-3d}. Alongside also the volume dependence of the propagator in the infrared and the behavior of it at zero momentum are shown. First of all, the statistically significant difference is quantitatively rather small, and confined to momenta below roughly half a GeV. The latter is expected, as the difference between both gauges is purely non-perturbative, and should therefore decrease as a power of momentum towards large momenta, leaving only the identical perturbative part of the propagators.

Interestingly, the statistically significant deviations lead to a gluon propagator which is smaller at low momenta than the one in minimal Landau gauge, in explicit agreement with condition \pref{propmin}. Any possibly existing deviations at larger momenta are not visible due to the statistical fluctuations\footnote{Note that the results for both gauges are obtained from two independent sets of configurations to reduce systematic errors. Hence a direct configuration-by-configuration comparison is not possible. Furthermore, fulfillment of the condition \pref{propmin} may be implemented at different momenta for different configurations.}.

Hence, the gluon propagator in absolute Landau gauge is more strongly infrared suppressed than in minimal Landau gauge. In particular, the gluon propagator at zero momentum decays faster in absolute Landau gauge than in minimal Landau gauge, and continues longer in a power-law decay towards the infinite-volume limit, while in minimal Landau gauge the gluon propagator seems to exhibit a mass-like behavior, as has been observed previously \cite{cucchieril7,Cucchieri:2007rg}. Only for the largest volume the value of the propagator at zero momentum seems to rise once more. However, as there is no similar effect for non-zero momentum, and the statistical error is comparatively large, this may, or may not, be a statistical fluctuation. Otherwise, it may be an effect due to the one observed in figure \ref{d-3d} that $\Delta$ decreases for large volumes. Beyond this, at least for the volumes investigated here, this decrease does not (yet) have an effect. In particular, excluding the largest volume, the value of $D(0)$ in minimal Landau gauge cannot be fitted with a pure power-law, while this is possible in the absolute Landau gauge. This is shown in figure \ref{gp-3d}. In general, the gluon propagator in absolute Landau gauge therefore could be vanishing at zero momentum in an infinite volume. In minimal Landau gauge, however, it seems to develop a mass, and becomes infrared constant. Furthermore, both behave similarly but not the same as a function of volume. However, larger volumes and better statistics will be necessary to substantiate this observation.

\begin{figure*}
\includegraphics[width=0.5\linewidth]{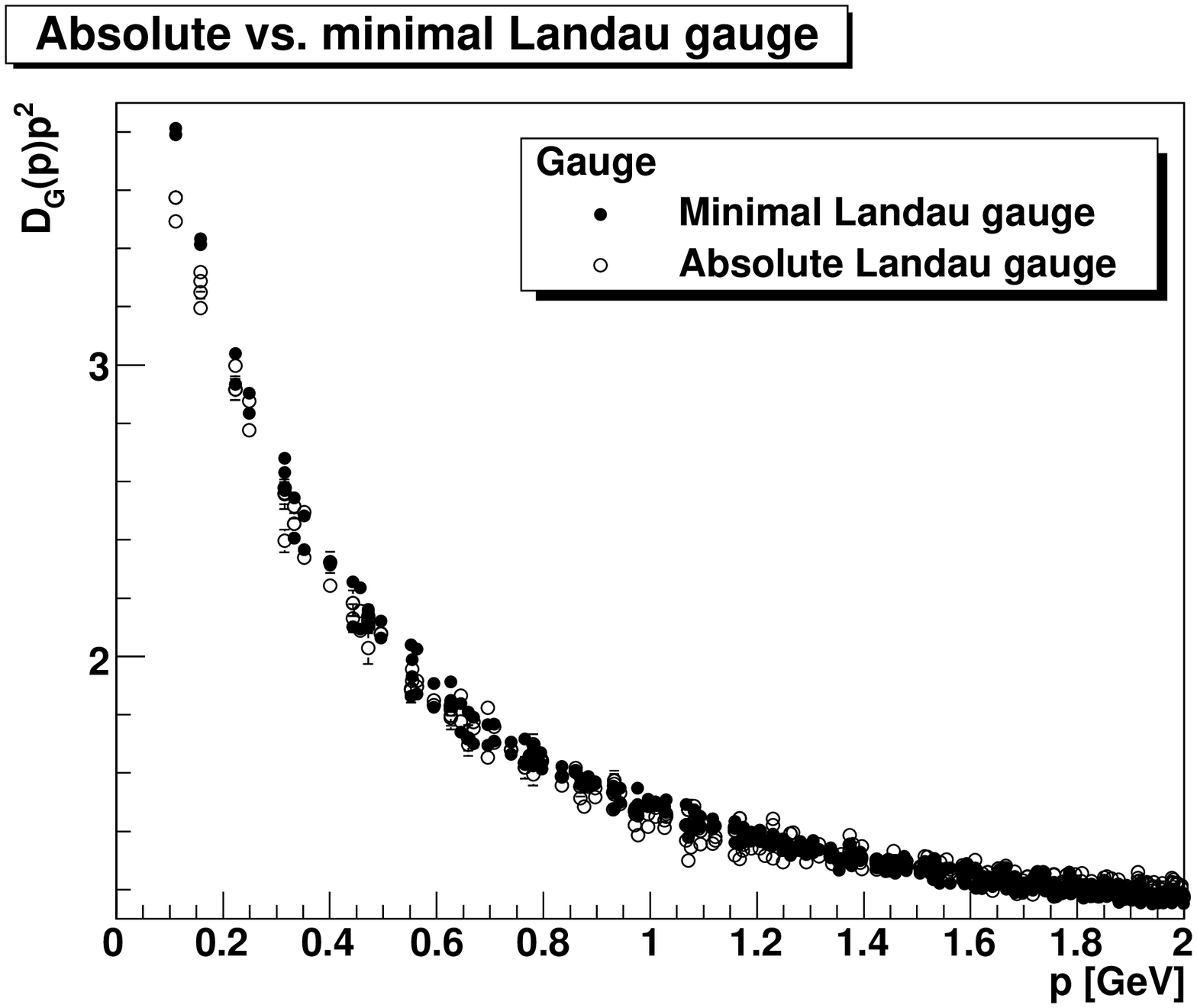}\includegraphics[width=0.5\linewidth]{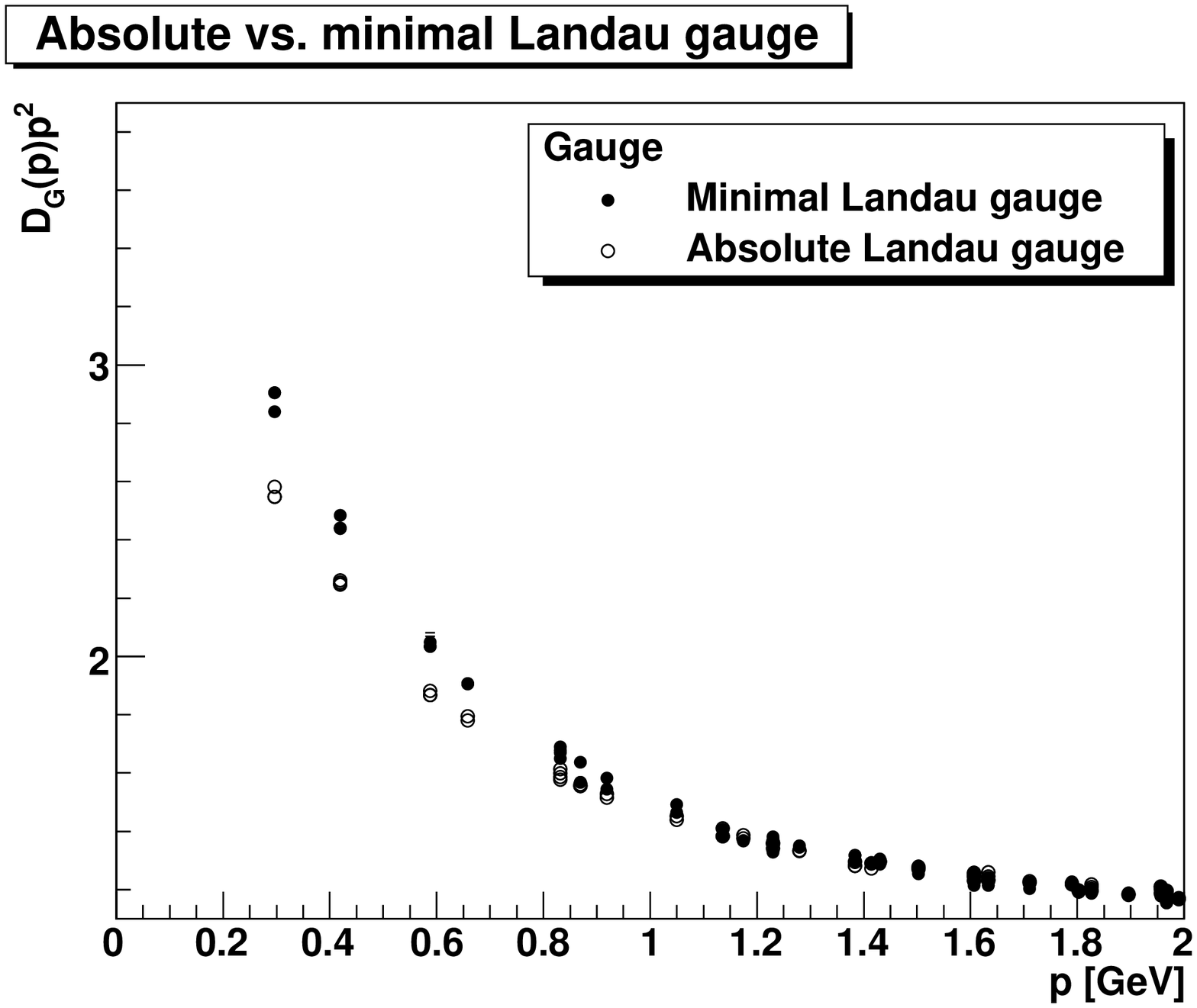}\\
\includegraphics[width=0.5\linewidth]{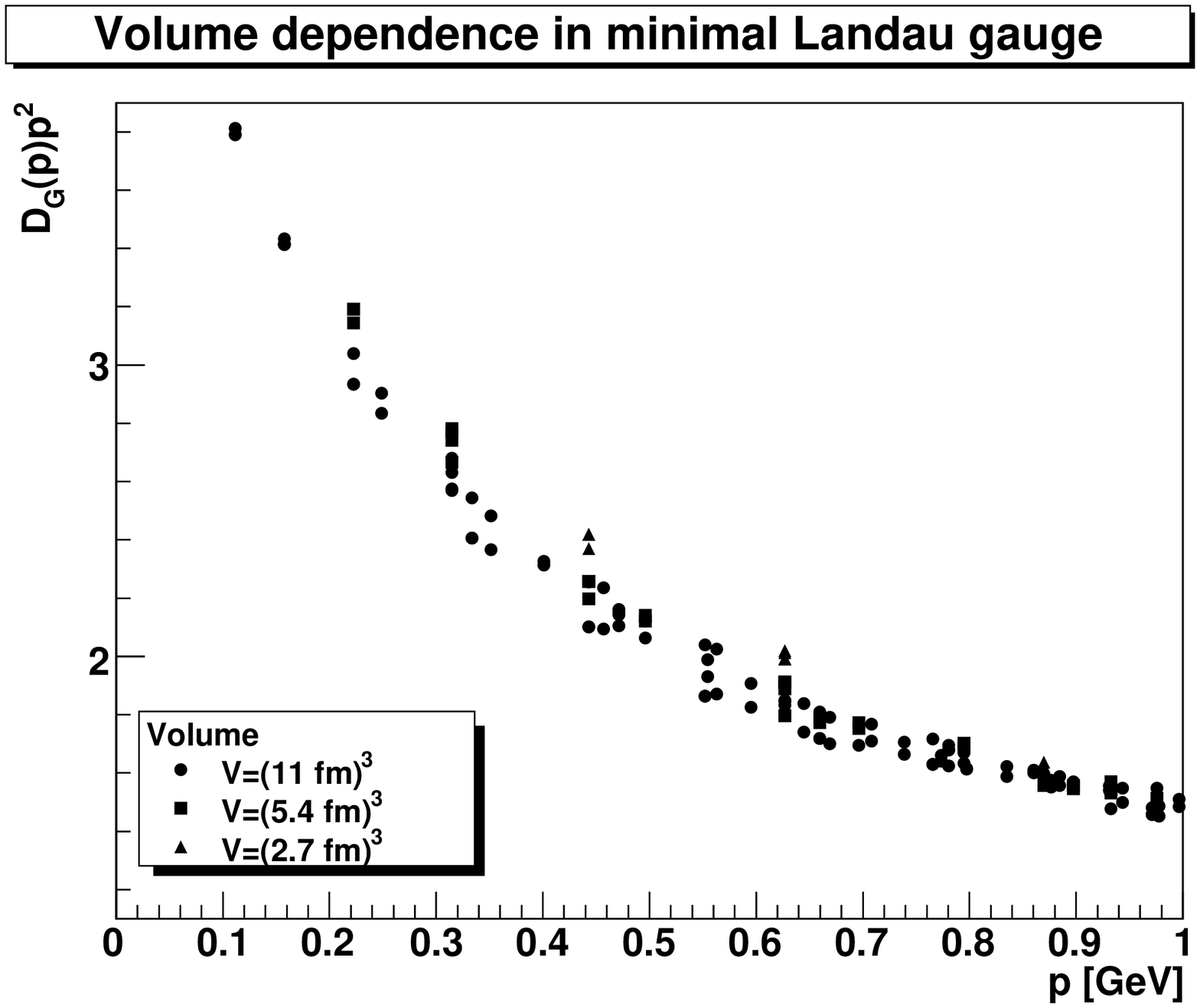}\includegraphics[width=0.5\linewidth]{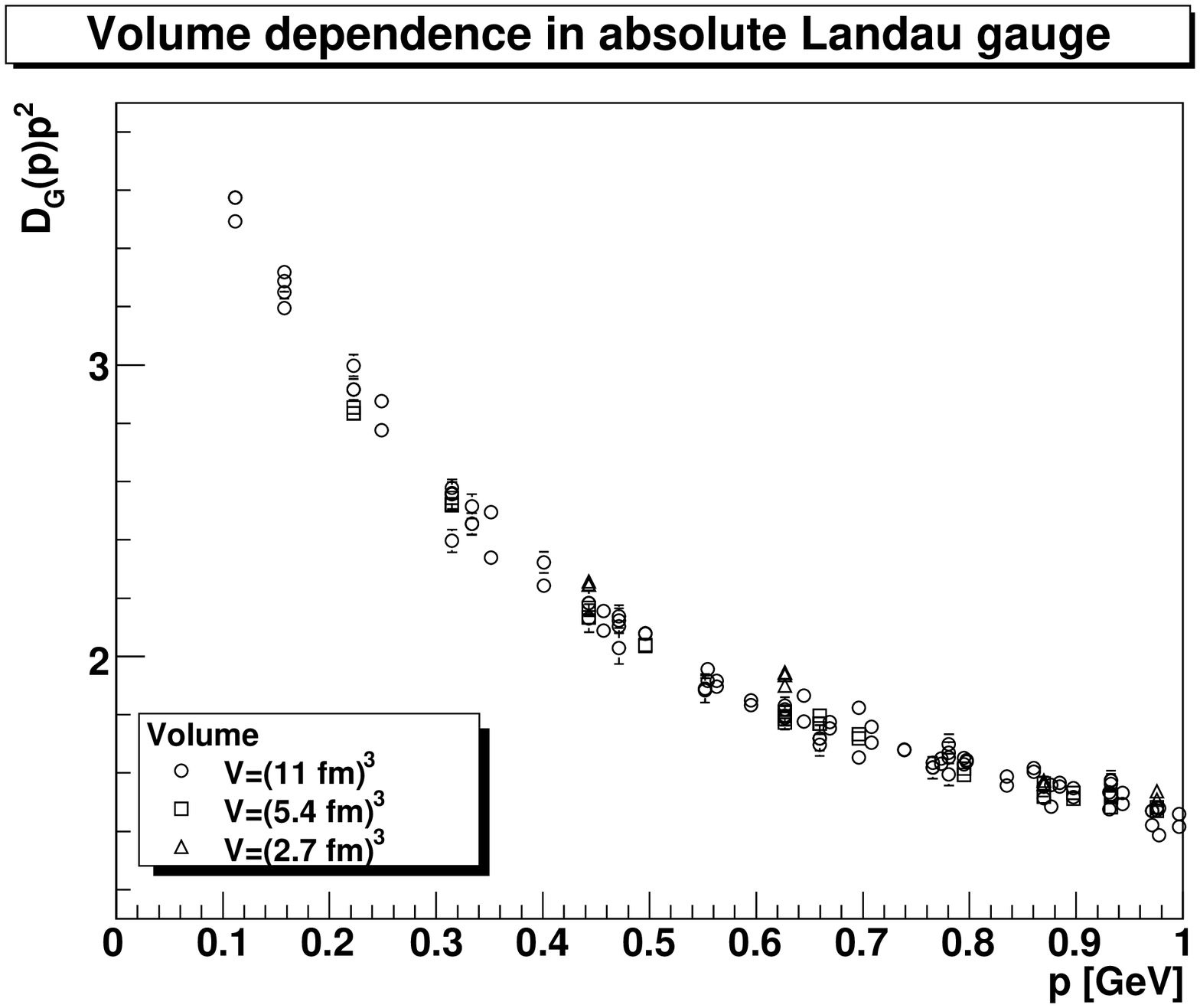}
\caption{\label{ghp-3d}The ghost dressing function in three dimensions. The top left panel shows the comparison between minimal (full circles) and absolute (open circles) Landau gauge on a $64^3$ lattice at $\beta=4.24$, corresponding to $V=(11$ fm$)^3$, the right panel shows the same comparison for a $24^3$ lattice, corresponding to $V=(4.1$ fm$)^3$. The bottom panels show the volume dependence for the ghost dressing function in minimal (left panel) and absolute (right panel) Landau gauge. Circles, squares, and triangles correspond to $V=(11$ fm$)^3$, (5.4 fm)$^3$, and (2.7 fm)$^3$, respectively, all calculated at $\beta=4.24$. }
\end{figure*}

Corresponding results for the ghost propagator are shown in figure \ref{ghp-3d}. The results are highly interesting. Comparing within a small volume the propagator in both gauges, it is found that the one in absolute Landau gauge is significantly smaller up to momenta as large as nearly 1 GeV than the one in minimal Landau gauge. In a larger volume, however, both look much more similar again. The reason for this behavior can be seen from the volume-dependence of the propagator in both gauges. In minimal Landau gauge, the propagator becomes less and less infrared enhanced with increasing volume, until it eventually becomes essentially the one of a massless particle \cite{cucchieril7,limghost}. This is, however, not the case in the absolute Landau gauge. While it also becomes less divergent for small volumes, the situation changes when Gribov copies become relevant. Then, the propagator stays at least as divergent as it was on smaller volumes, or may possibly become even more divergent again.

These results are reflected also in the spectrum of the Faddeev-Popov operator, in particular on its lowest eigenvalue. However, the statistical uncertainty in this case is too large for a definite conclusion.

\begin{figure}
\includegraphics[width=\linewidth]{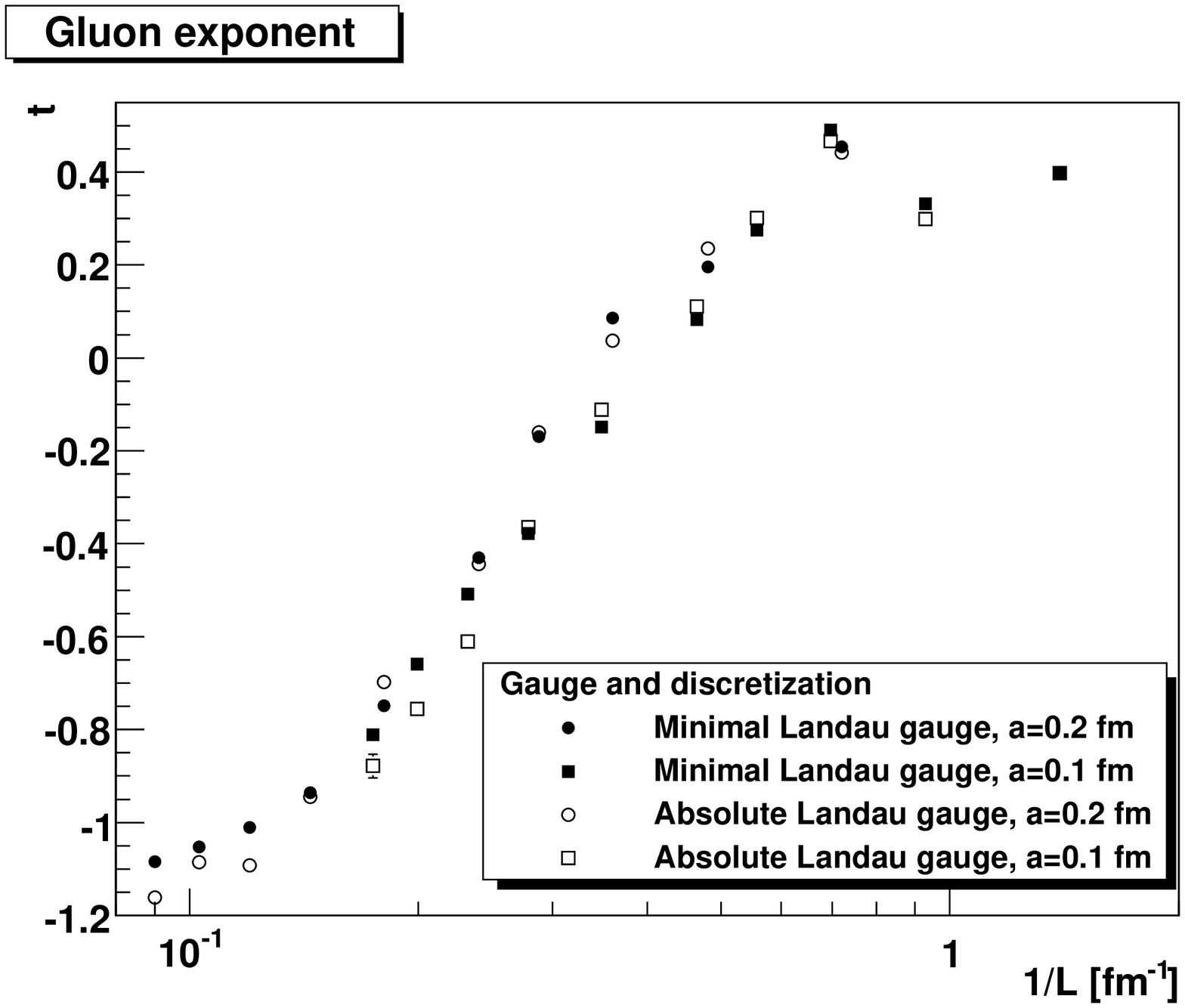}\\
\includegraphics[width=\linewidth]{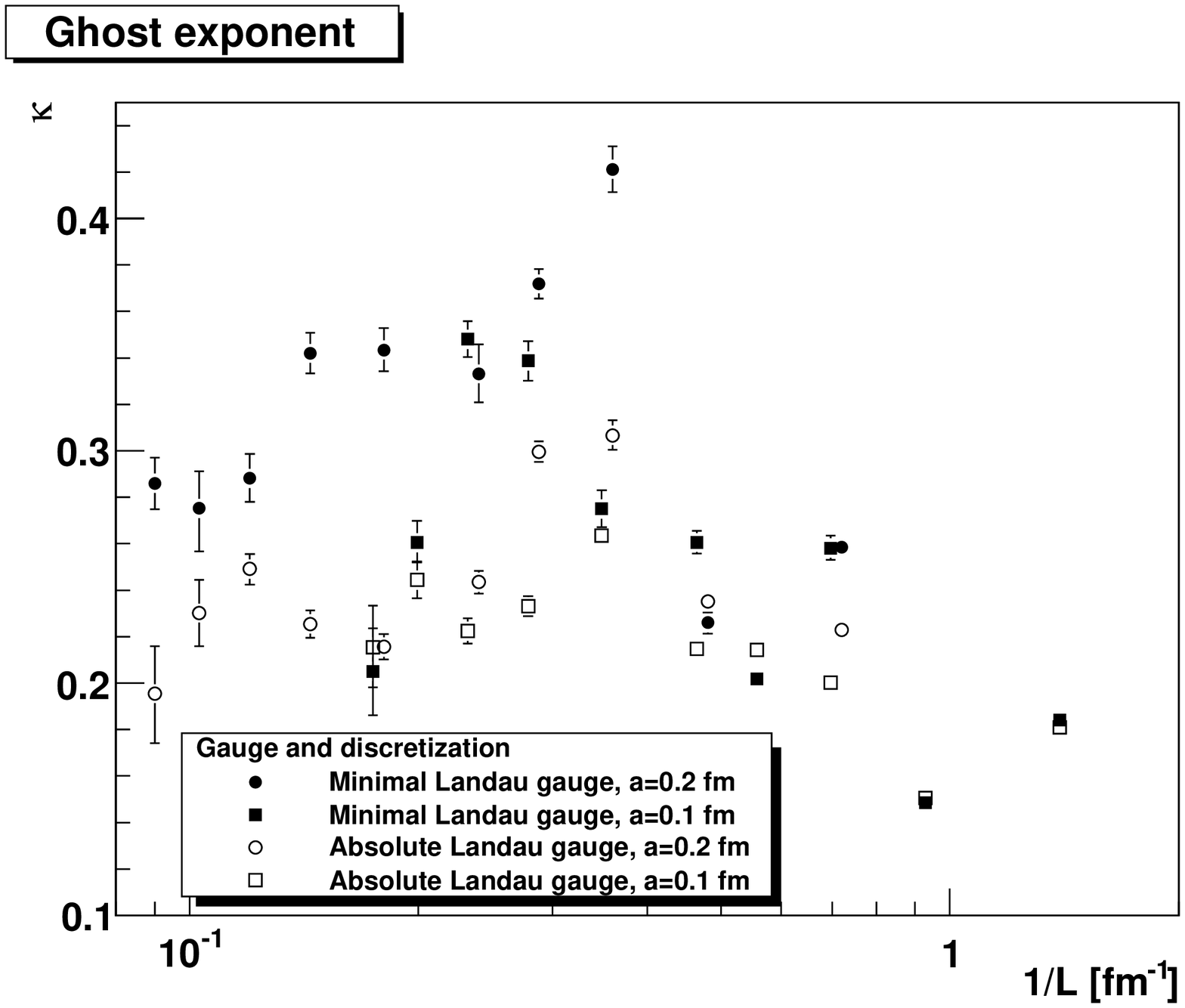}
\caption{\label{d3-exp}The gluon exponent $t$ (top panel) and the ghost exponent $\kappa$ (bottom panel) as a function of volume in three dimensions. Full and open symbols correspond to the minimal and absolute Landau gauge, respectively. Circles are from calculations at $\beta=4.24$ and squares from calculations at $\beta=7.09$.}
\end{figure}

To complete the collection of results on the propagators, two more properties are investigated. One is again the differential properties in the form of the effective exponents $t$ and $\kappa$, which are shown in figure \ref{d3-exp}. The exponent for the gluon propagator is mildly affected at large volumes, becoming more negative in the absolute Landau gauge. Still, the volumes employed here are such that also in minimal Landau gauge the exponent is smaller than $-1$, but larger than the one in absolute Landau gauge. Both are thus for these volumes in accordance with an infrared vanishing gluon propagator. Of course, this is no longer the case on larger volumes in case of the minimal Landau gauge \cite{cucchieril7}, but it is currently unknown what will be the case for the absolute Landau gauge. In case of the ghost exponent, the difference between both gauges is more pronounced, but statistical uncertainties prevent a definite conclusion at large volumes. In particular, as can already be deduced from the ghost dressing function in figure \ref{ghp-3d}, the ghost exponent is initially larger in minimal Landau gauge than in absolute Landau gauge. The exponent in minimal Landau gauge, however, decreases with volume  \cite{cucchieril7,limghost} while the one in the absolute Landau gauge is essentially volume independent within statistical accuracy. This once more demonstrates that the results in absolute Landau gauge exhibit different finite volume effects.

\begin{figure}
\includegraphics[width=\linewidth]{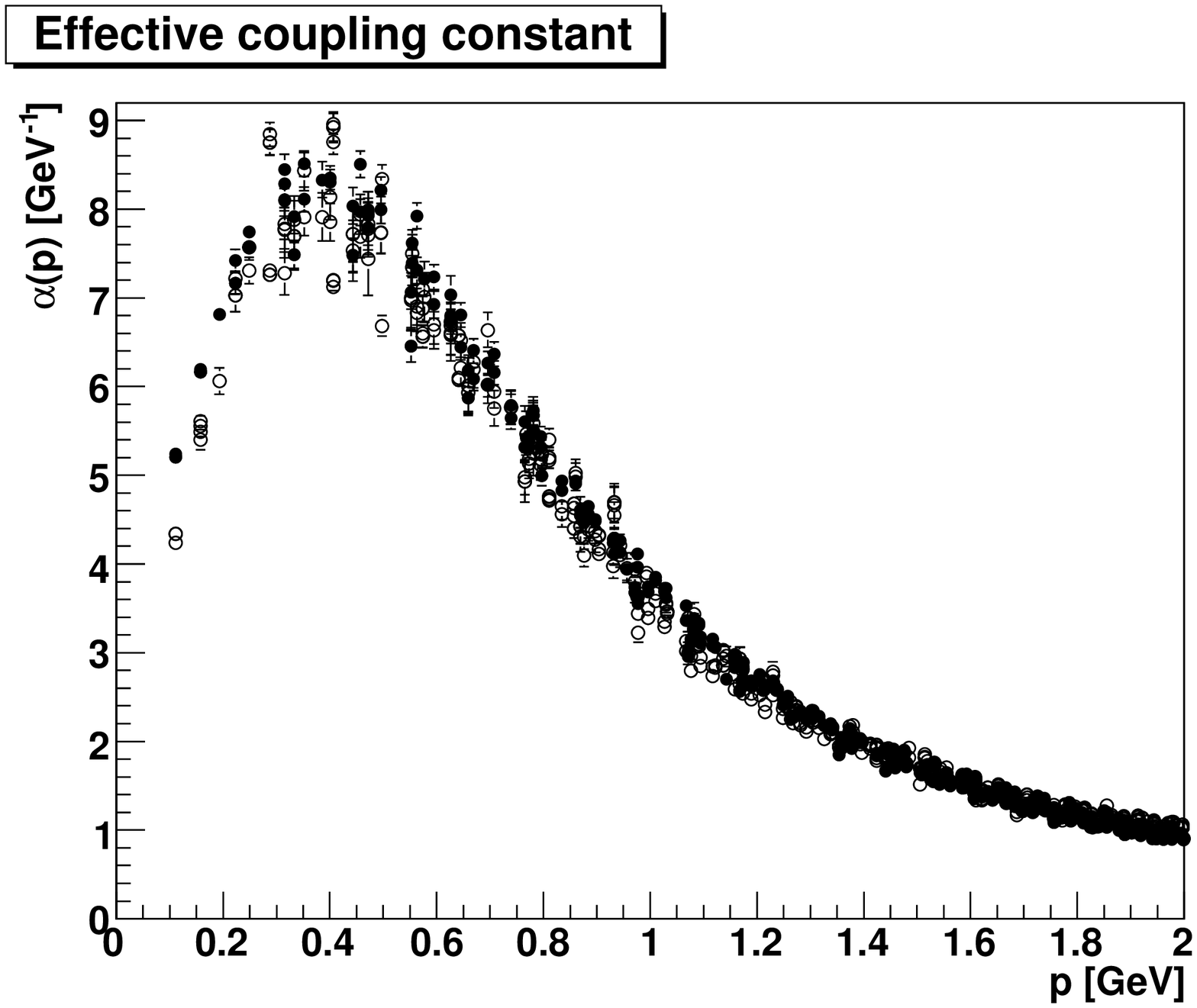}\\
\includegraphics[width=\linewidth]{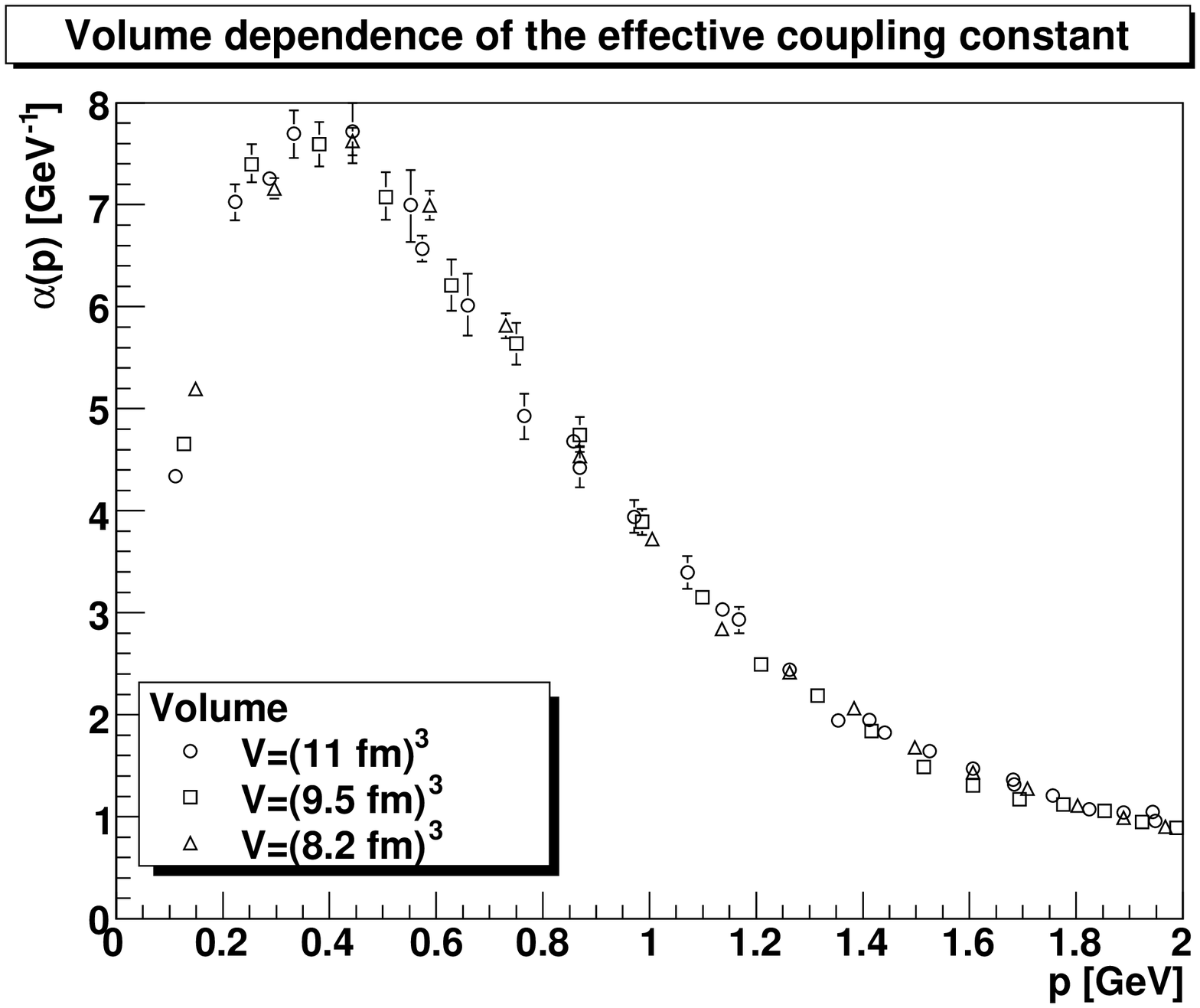}
\caption{\label{d3-alpha}In the top panel the quantity $\alpha$, defined in \pref{alpha} is shown in minimal Landau gauge (full circles) and in absolute Landau gauge (open circles). In the lower panel the volume dependence of the effective coupling constant in absolute Landau gauge is shown. Circles are from a $64^3$ lattice, squares from a $56^3$ lattice, and triangles from a $48^3$ lattice, all calculated at $\beta=4.24$.}
\end{figure}

Finally, the quantity $\alpha$, defined in equation \pref{alpha}, should become constant for all dimensions in the far infrared, and thus also in three dimensions. This is the case in two dimensions in both gauges, as shown above. However, it seems to go to zero in three and higher dimensions \cite{cucchieril7}. This is to be expected in any finite volume \cite{Fischer:2007pf}, but the pattern of finite volume artifacts observed in minimal Landau gauge is not in agreement with a constant value in the infinite volume limit \cite{Cucchieri:2007rg,limghost}. The result in absolute Landau gauge, and its volume dependence, is shown in figure \ref{d3-alpha}. The difference between both gauges is not qualitative, in particular, in both cases $\alpha$ seems to go to zero. However, there is an interesting volume dependence observed in the absolute Landau gauge. It is seen that with increasing volume the maximum tends to move towards smaller momenta, albeit rather slowly. The consequence of this is particularly visible in case of the second-lowest momentum point: The same value of $\alpha$ is attained at smaller and smaller momenta on the decreasing slope. This would be in accordance with the type of finite volume effects predicted for an infrared finite coupling constant in the infinite-volume limit \cite{Fischer:2007pf}.

\subsection{A note on four dimensions}

\begin{figure}
\includegraphics[width=\linewidth]{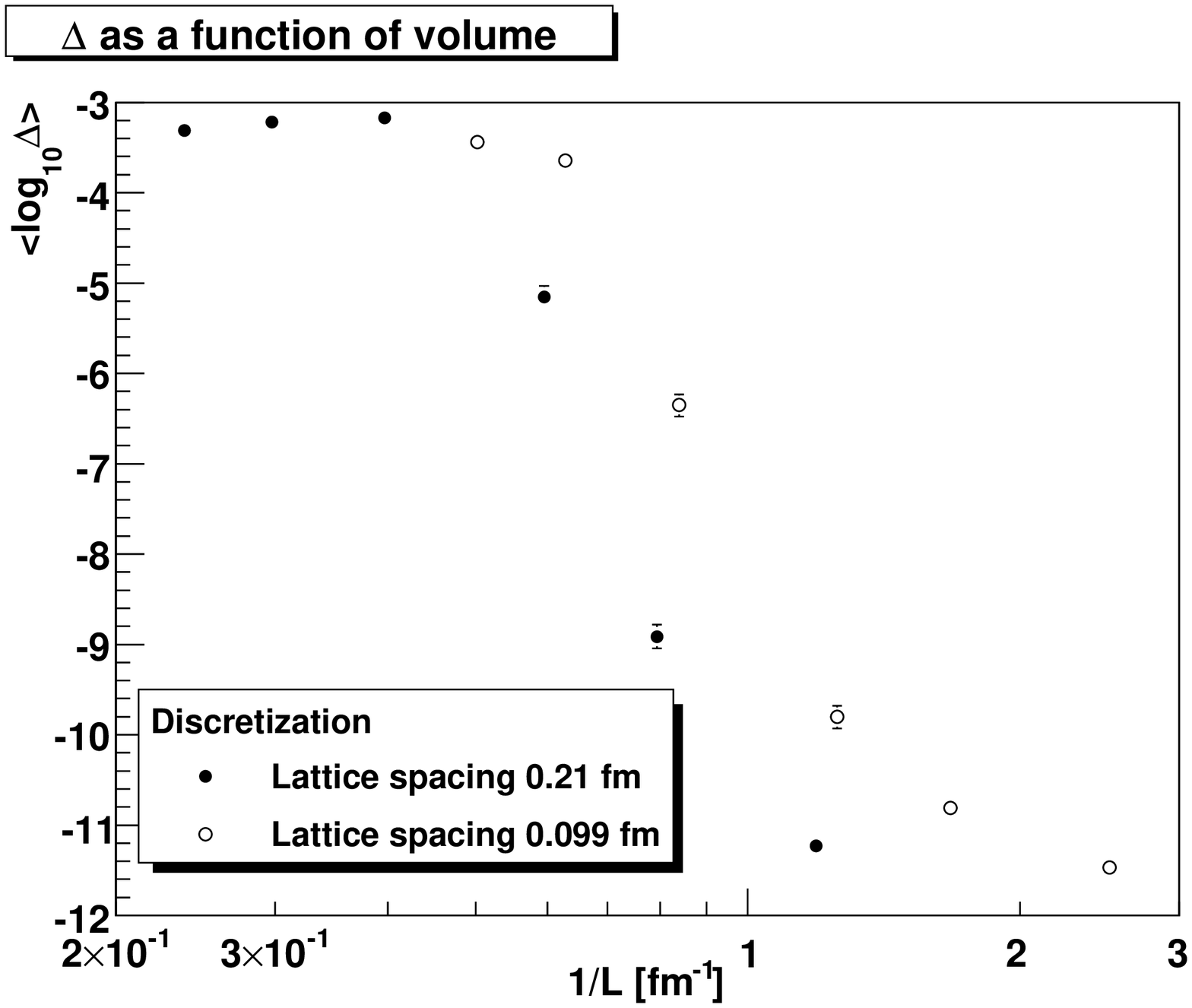}
\caption{\label{ld-4d}The presence of Gribov copies as measured by $\Delta$ as a function of inverse volume in four dimensions. Full circles correspond to $\beta=2.2$, while open circles correspond to $\beta=2.45$.}
\end{figure}

Take now for a moment for granted that in absolute Landau gauge the infrared behavior of the Green's functions would be given by power laws. This would imply that the effective gluon exponent would become smaller with increasing dimension, while the one for the ghost becomes larger \cite{gzwanziger}. Given the volumes from which onward a maximum in the gluon propagator can be observed in two and three dimensions, it can be immediately deduced that lattice volumes of the order of $50^4$ at $\beta=2.2$ would be necessary to obtain such a maximum unambiguously in four dimensions. Furthermore, from the results for $\Delta$ obtained on small lattices, which is shown in figure \ref{ld-4d} and should be compared to the figures \ref{ld-2d} and \ref{ld-3d}, it is clear that the number of Gribov copies in four dimensions at a given linear lattice extension in lattice units significantly exceeds the same number in lower dimensions. As assured in the beginning the issue of Gribov copies becomes more severe with an increasing number of dimensions.

From this already a rough estimate of the necessary computing time can be deduced, which will be needed to obtain the propagators in absolute Landau gauge for the indicated volume. This exceeds the amount of computing time invested here for the the calculations in two and three dimensions by more than an order of magnitude. Aside from this, the necessary amount of storage space at intermediate steps of the calculations to hold all the gauge transformations needed in parallel for the algorithm presented in appendix \ref{sabslan} is also rather prohibitive. Hence, this has to be investigated in detail further, when sufficient resources will be available in the near future.

\section{Discussion}\label{sdiscuss}

\subsection{Non-perturbative definitions of the Landau gauge}\label{ssgauge}

Within the first Gribov region many Gribov copies exist. All of the gauge-fixing techniques used in this work select for each configuration one particular of these minima. However, which of these minima is selected depends on the algorithm. Two different versions which use an implicit probabilistic choice are the adaptive stochastic overrelaxation procedure, which has been used as defining the minimal Landau gauge throughout, with and without a limit on the number of gauge-fixing sweeps. The probabilistic aspect comes into play as the Gribov copy chosen depends on the initial (random) guess for the correct gauge transformation to Landau gauge. In this sense, both lead to different average values of the gauge-fixing functional ${\cal F}(A)$ \pref{fumod}, defined by their different weighting functions. Although the differences on the level of Green's functions are small, there exist differences for both possibilities, as shown explicitly in section \ref{sgf}.

An alternative definition is by the selection of the lowest minimum (found). This is the method used for the absolute Landau gauge. Again, this corresponds to a different value of ${\cal F}(A)$, and also to different Green's functions, as has been shown in section \ref{sgauge}. It would, of course, be possible to select instead of the lowest minimum, the second-lowest, the third-lowest etc., which gives a one-parameter-family of Landau gauges\footnote{On a finite lattice, there will be no degeneracies. In the infinite-volume and continuum limit, this may be different, and a prescription how to deal with these degeneracies would be required as well.}. Of course, all of them, by construction, lead to a different value of ${\cal F}(A)$, and therefore, by virtue of \pref{propmin}, to differing gluon propagators.

This, so far, results from the possibilities how to define the Landau gauge, and are consequences of the non-perturbative ambiguities of the Landau gauge condition \pref{landau}.

\subsection{A possible interpretation}\label{ssint}

It is now possible to interpret these concepts and results using an existing framework \cite{dse}. As the data is not yet yielding unambiguously one result, given the limited resources available, this is, of course, just a possibility.

It is conceivable that gauges based on selecting a particular minimum out of the list of ordered minima will yield Green's functions with values between the results in the minimal Landau gauge and the absolute Landau gauge, as long as the respective values of ${\cal F}(A)$ are between those of the absolute and the minimal Landau gauge. In case of values larger than the one of minimal Landau gauge, it is not a-priori clear what might occur. Although this conjecture is likely, it requires explicit future proof.

In particular, it would be interesting to track whether the value of $D(0)$ or $p^2D_G(p)|_{p=0}$ could be mapped to such a definition of the non-perturbative Landau gauge. If this would be the case, it is tempting to assume that this family of Landau gauges could be in one-to-one correspondence to the one-parameter family of solutions found for the corresponding functional equations \cite{dse}. Furthermore, this would give a direct explanation why the corresponding effective action in minimal Landau gauge is not having a unique minimum (to one-loop order) as a (implicit) function of $D(0)$ \cite{Dudal:2008rm}, at least in three dimensions: The value of $D(0)$ (or better $p^2D_G(p)|_{p=0}$ \cite{dse}) would then have no physical significance, but only acts as a gauge parameter. This would be, of course, highly welcome, as $D(0)$ cannot be an independent parameter of the theory, which is completely specified once the coupling, or the scale, has been fixed. On the lattice, of course, this family is finite, as only a finite number of Gribov copies exists (or at least are found), while this is a continuous one-parameter family in the continuum.

A possibility would then be that the absolute Landau gauge in fact corresponds to the endpoint of this one-parameter family, i.\ e., to the value $(p^2D_G(p))^{-1}|_{p=0}=0$. Whether in the continuum the scaling solution possibly fulfills the condition \pref{propmin} for the absolute Landau gauge, and to which extent this affects solutions with a finite value for $(p^2D_G(p))^{-1}|_{p=0}$ is discussed in \cite{dse}. However, these are currently truncation-dependent statements \cite{dse}.

Also the results obtained in Landau gauge here exhibit a dependence on volume, which would be compatible with the one expected for a power-law solution in the infinite-volume and continuum limit \cite{Fischer:2007pf}, although the results are not unambiguously in favor of this, they are neither in strict disagreement. In particular, all changes which occur when moving from the minimal Landau gauge to the absolute Landau gauge are exactly as expected from the scenario sketched here.

The only, at first sight, disagreeing result is the ghost propagator in absolute Landau gauge being smaller than in minimal Landau gauge. The reason for this behavior of the ghost propagator, shown in figure \ref{ghp-3d} in three dimensions, may be connected to the volume evolution of the first Gribov region and the fundamental modular region. Due to volume effects, in both cases configurations at the boundary dominate the average. However, at any finite volume both do not have a common boundary \cite{gzwanziger,gzwanziger2}. Given that the first Gribov region is bounded by configurations with the smallest eigenvalues of the Faddeev-Popov operator, it appears that the ghost propagator has to diverge less in absolute Landau gauge than in minimal Landau gauge \cite{Cucchieri:1997dx,limghost}. But not only the eigenvalues enter the determination of the ghost propagator, but also the eigenstates \cite{Sternbeck:2005vs}. Although occasionally assumed, the Faddeev-Popov operator does not have in the infinite-volume and continuum limit only normalizable eigenstates. This is due to its equivalence to a quantum mechanical Hamilton operator, which can have scattering states as eigenstates. In particular, explicit examples exist for cases where the lowest (zero) eigenvalue has a non-normalizable eigenstate \cite{Maas:2005qt}. This is also an important fact in the Gribov-Zwanziger framework \cite{gzwanziger,gzwanziger2}. On the other hand, in any finite volume, the eigenstates can be normalized, and taking the (possible) divergence of this normalization in the infinite-volume limit into account could be crucial. The combined effect of the precise shape of the low-eigenvalue spectrum and a diverging normalization of the eigenstates could be sufficient to provide a more infrared divergent ghost propagator in the infinite-volume and continuum limit in absolute Landau gauge than in minimal Landau gauge. Identifying how this could actually be realized is, however, beyond the scope of this work. A final proof is, of course, only that in the absolute Landau gauge at sufficiently large volume the ghost propagator would be more singular than in the minimal Landau gauge. The volume-dependence of the propagator in both gauges found here is as expected if this is the case.

A consequence of this scenario is that it should be expected that also in two dimensions, for sufficiently large volumes and number of Gribov copies, an infrared finite gluon propagator is obtained in the minimal Landau gauge. Given the results shown in figure \ref{ld-2d}, this must be at volumes much larger than the ones currently investigated. This will be checked in the future. Corresponding observations have already been made analytically in Coulomb gauge in two dimensions \cite{Reinhardt:2008ij}.

Finally, the question of discretizations artifacts has so far not been sufficiently addressed, although indications exist that these might still be relevant \cite{lvs}. This will also be investigated in the future. In other gauges, e.\ g.\ linear covariant ones, they are of considerable importance \cite{rank,Cucchieri:2008zx}.

In addition, from the lattice perspective it is not fully solved how this relates to the Kugo-Ojima scenario, as this requires the construction of a lattice version of the BRST symmetry, which is currently under investigation \cite{vonSmekal:2008en}. Given such a construction and further assumptions, a power-law behavior of the Green's functions is implied  \cite{dse}. This global BRST, of course, is not necessarily simply related to the ordinary perturbative BRST symmetry outside the domain of applicability of perturbation theory, and therefore statements on just the perturbative BRST symmetry cannot be carried over to the non-perturbative domain\footnote{E.\ g., the symmetry may not be manifest, as Lorentz symmetry at finite temperature. Alternatively, the global BRST symmetry may be well-defined and unbroken, while the ordinary one would not be, similar to translation symmetry on a sphere.}. It should be noted that in the Kugo-Ojima construction it is actually not relevant which particular non-perturbative extension of the Landau (or linear covariant) gauge is used. The only ingredient necessary is the existence of an unbroken BRST charge and its algebraic properties, as can be seen directly from the construction \cite{Kugo}. The existence of Gribov copies then implies that such a charge has necessarily to be constructed globally. Conversely, this implicitly defines a unique, non-perturbative gauge.  However, an explicit construction would be necessary, or at least a proof of its existence. Until then, the existence of such a charge can only be assumed, as has been done in \cite{dse}. Whether this gauge is then identical to the absolute Landau gauge can be checked, at least in principle, by the condition \pref{fumod}. An infrared singular ghost dressing function in the absolute Landau gauge would, however, be already an indication of such an equivalence.

This completes the scenario, which would provide a picture where both, decoupling-type solutions and scaling-type solutions occur, though in different gauges. Of course, this is only one possible scenario.

\begin{figure}
\includegraphics[width=\linewidth]{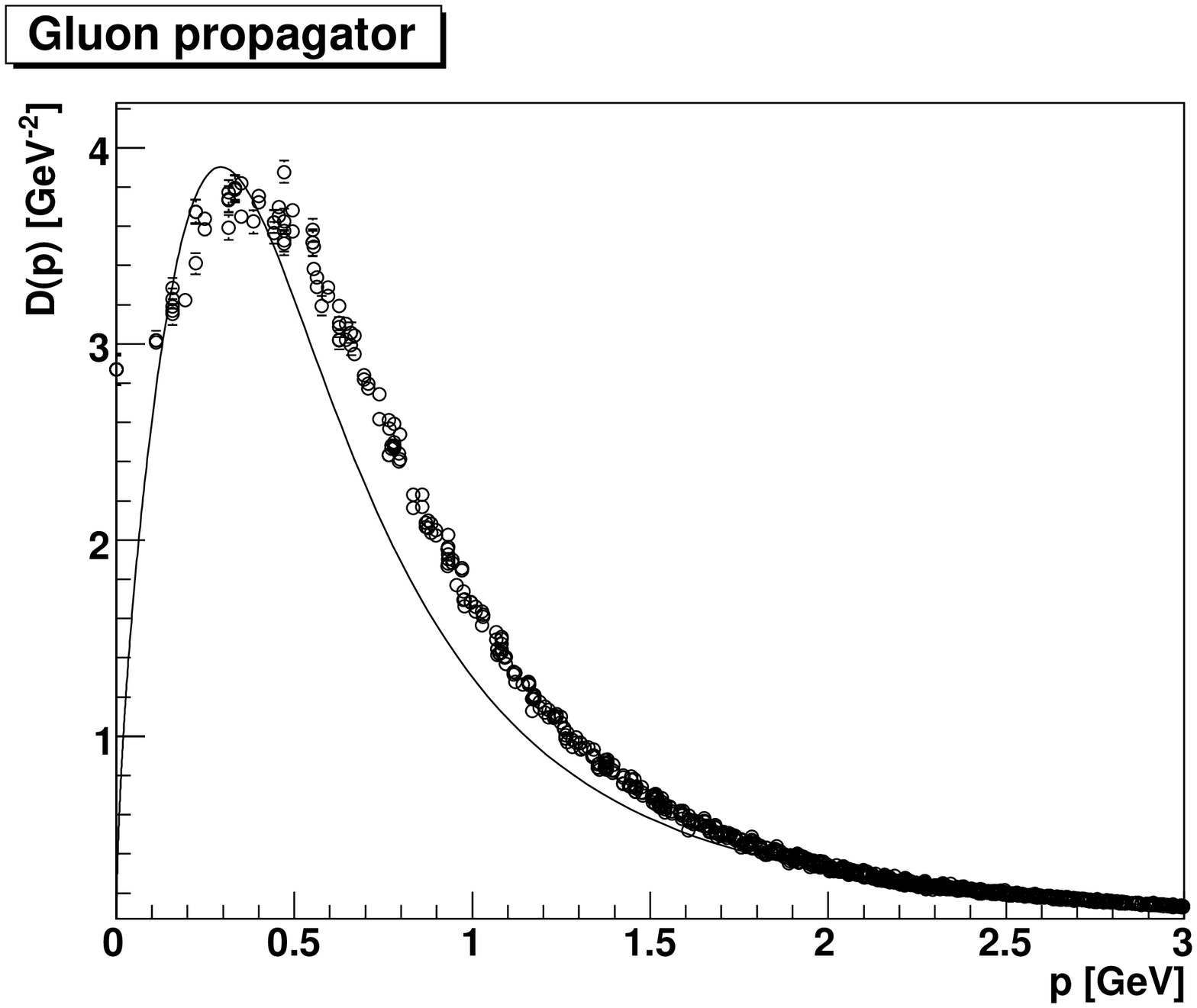}\\
\includegraphics[width=\linewidth]{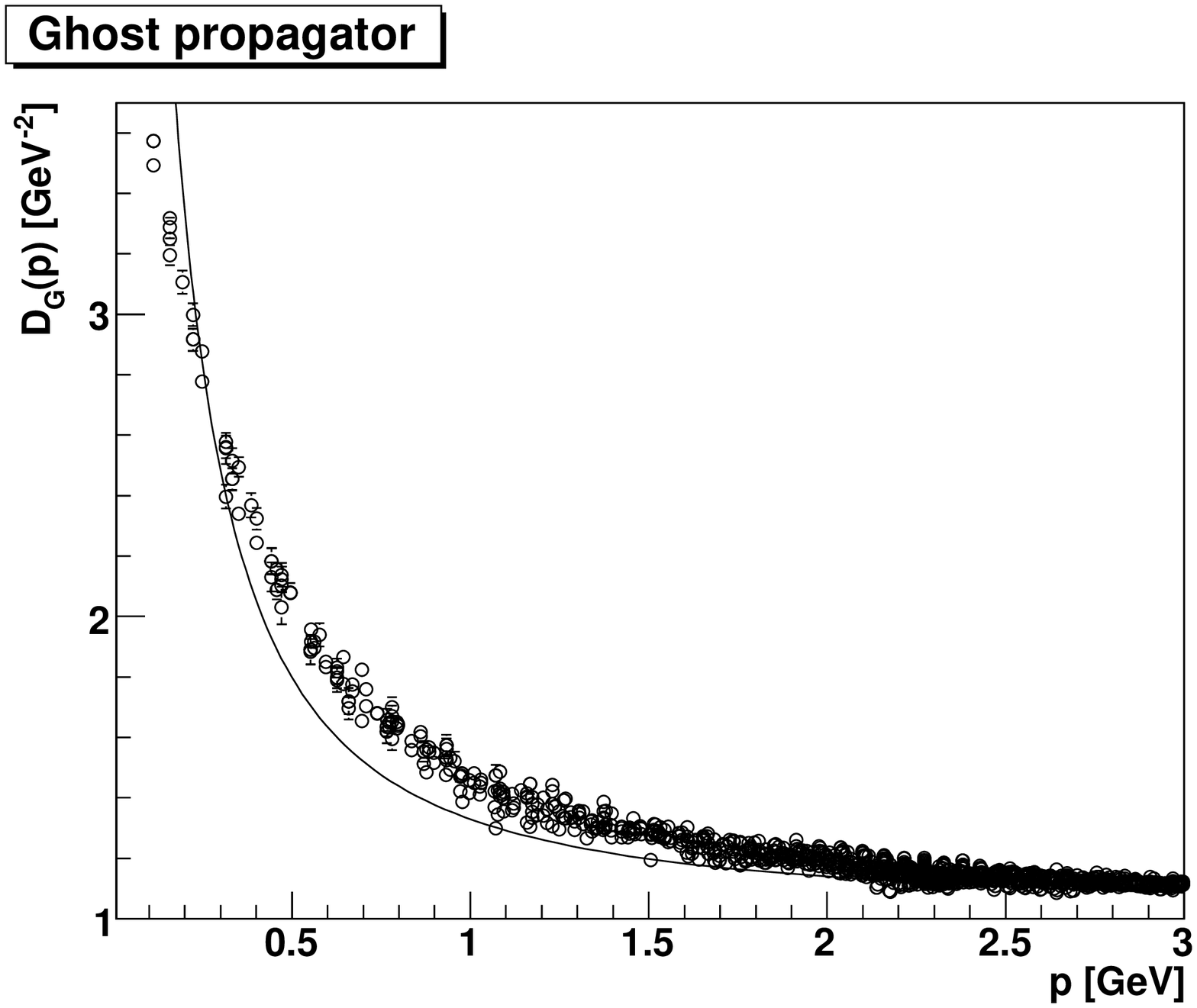}\\
\caption{\label{dse-comp}The lattice results in absolute Landau gauge from a $64^3$ lattice at $\beta=4.24$ compared to results from DSE calculations \cite{Maas:2004se}.}
\end{figure}

Hence, at the current point it seems more appropriate to compare the lattice results in absolute Landau gauge, rather than in minimal Landau gauge, with those from functional results which exhibit a scaling behavior in the far infrared. This is shown in figure \ref{dse-comp}. The agreement is rather good at these small volumes.

\section{Summary}\label{ssum}

When performing gauge-fixed calculations in the non-perturbative regime, it is necessary to precisely specify how to treat the residual gauge orbit which exists due to Gribov-Singer copies. In particular, gauge-dependent correlation functions like propagators can depend on how the residual gauge orbit is treated.

The results here show that this is indeed the case when comparing the propagators in the minimal Landau gauge and the absolute Landau gauge. The differences quickly diminish with momenta, and are quantitatively essentially irrelevant at momenta larger than 1 GeV. However, in the far infrared region they lead to different behaviors. Nonetheless, the changes are compatible with even a qualitative change of the infrared properties, although with the currently available statistics and volumes a final conclusion cannot be drawn, and the effects of discretization effects is to some extent unexplored \cite{lvs}. It has also be demonstrated that the effects at a fixed volume become more severe with an increasing number of dimensions.

This is only the result of the comparison of two of the possible non-perturbative completions of the perturbative Landau gauge. As discussed, at least on the lattice it is possible to construct a whole one-parameter family of Landau gauges all corresponding to the perturbative Landau gauge. A possible scenario of how these may be related to the one-parameter family of solutions found in the corresponding functional calculations in the infinite-volume and continuum limit \cite{dse} has been presented in section \ref{ssint}.

There results emphasize that in the non-perturbative domain it is necessary to really completely specify the gauge-fixing procedure, also in functional methods, as the different gauges can correspond to different Green's functions. Comparisons between different methods can only be made if the same gauge has been employed.

\acknowledgments
 
I am grateful to R.~Alkofer, A.~Cucchieri, C.~S.~Fischer, H.~Gies, T.~Mendes, J.~M.~Pawlowski and L.~von~Smekal for helpful discussions. This work was supported by the DFG under grant number MA 3935/1-2 and by the FWF under grant number P20330. Part of the computing time was provided by the Slovak Grant Agency for Science, grant VEGA No.\ 2/6068/2006, and by the Helmholtz association, grant No.\ VH-NG-332 (Helmholtz Young Investigator group ``Nonperturbative Phenomena in QCD''). I am indebted to C.~S.~Fischer and J.~M.~Pawlowski for a critical reading of the manuscript and helpful comments. The ROOT framework \cite{Brun:1997pa} has been used in this project.

\appendix

\section{Technical details of the lattice simulations}\label{stech}

The configurations have been obtained using a hybrid heat-bath overrelaxation procedure using a standard Wilson action. Details of this procedure and the determination of the propagators along with statistical error determination procedures can be found in \cite{Cucchieri:2006tf}.

\begin{table}
\caption{\label{tcgf}
Number of configurations considered in the numerical simulations for the results of section \ref{sgf}. The value of the lattice spacing $a$ has been taken from Ref.\ \cite{Cucchieri:2003di}. All calculations have been performed in three dimensions on a 40$^3$ lattice at $\beta=4.2$. Always 200 thermalization sweeps and 50 sweeps between two consecutive measurements have been performed. 'Limit' is the maximum number of permitted gauge fixing sweeps, and 'Conf.' the number of configurations used. The limit $\infty$ corresponds to no restriction on the number of gauge-fixing sweeps, and the number in parentheses is the actual maximum number observed. The data in this case is taken from \cite{Cucchieri:2008qm}, and the first and second number of configurations refers to the number of configurations used for the ghost and gluon propagator, respectively. Note that for each limit case independent configurations have been obtained. Note further that in all cases multiple independent runs (about 15-20) have been performed to obtain the data.}
\begin{ruledtabular}
\vspace{1mm}
\begin{tabular}{|c|c||c|c||c|c|}
Limit & Conf. & Limit & Conf. & Limit & Conf. \cr
\hline
$\infty$ (28740) & 1077/8175 & 27303 & 1091 & 25938 & 1262 \cr
\hline
24641 & 1030 & 23409 & 1021 & 22239 & 1029 \cr
\hline
21127 & 1016 & 20071 & 1060 & 19067 & 1148 \cr
\hline
18114 & 1043 & 17208 & 1088 & 16348 & 1150 \cr
\hline
15531 & 1080 & 14754 & 1083 & 14016 & 1004 \cr
\hline
13315 & 1154 & 12649 & 1125 & 12017 & 1106 \cr
\hline
11416 & 1135 & 10845 & 1032 & 10303 & 1032 \cr
\hline
9788 & 1128 & 9299 & 1134 & 8834 & 1154 \cr
\hline
8392 & 1154 & 7972 & 1050 & 7573 & 1154 \cr
\hline
7195 & 1154 & 6835 & 1131 & 6493 & 1184 \cr
\hline
6168 & 1131 & 5860 & 1213 & 5567 & 1088 \cr
\hline
5289 & 1116 & 5025 & 1161 & 4773 & 1047 \cr
\hline
4535 & 1237 & 4308 & 1150 & 4093 & 1131 \cr
\hline
3888 & 1128 & 3694 & 1138 & 3509 & 1136 \cr
\hline
3334 & 1158 & 3167 & 1131 & 3009 & 1154 \cr
\hline
2858 & 1154 & 2715 & 1048 & 2579 & 1112 \cr
\hline
2450 & 1111 & 2328 & 1113 & 2212 & 1103 \cr
\hline
2101 & 1078 & 1996 & 1061 & 1896 & 1153 \cr
\hline
1801 & 1153 & 1711 & 967  & 1626 & 967  \cr
\hline
1545 & 986  & 1467 & 1026 & 1394 & 991  \cr
\hline
1324 & 1088 & 1258 & 853  & 1195 & 752  \cr
\hline
1135 & 593  & 1078 & 1136 & 1024 & 854  \cr
\hline
972  & 666  & 923  & 554  & 876  & 781  \cr
\hline
832  & 407  & 791  & 632  & 751  & 366  \cr
\hline
714  & 80   & 678  & 196  & 644  & 148  \cr
\hline
\end{tabular}
\end{ruledtabular}
\end{table}

In total, three sets of data have been obtained. One set has been used in section \ref{sgf} to investigate properties of the adaptive stochastic overrelaxation gauge-fixing procedure \cite{Cucchieri:2006tf}, which is used as an element in the method to fix to the absolute Landau gauge described in the next appendix \ref{sabslan}. The details of these simulations can be found in table \ref{tcgf}. In this set of data, the gauge-fixing procedure was limited in the number of iterations, as described in section \ref{sgf}. For each limit, an independent set of configurations was used, to reduce systematic errors.

Note that for all data sets in tables \ref{tcgf}-\ref{talg4} multiple independent runs were always performed to reduce auto-correlation effects. In case of the largest volumes, usually less than ten configurations constitute an independent run.

\begin{table}
\caption{\label{tmlg2}
Number of configurations considered in the numerical simulations for the results in minimal Landau gauge of section \ref{sgauge} in two dimensions. The value of the lattice spacing $a$ has been determined analytically \cite{Dosch:1978jt}. To assign units the standard value of (440 MeV)$^2$ has been assigned to the string tension. $N$ is the number of lattice sites in one direction of the symmetric lattice, $a$ is the lattice spacing, $L$ the physical length extent, 'Therm.' the number of thermalization sweeps,  'Sweep' the number of sweeps between two consecutive measurements and 'Conf.' the total number of configurations. Note that in all cases multiple independent runs have been performed to obtain the data.}
\begin{ruledtabular}
\vspace{1mm}
\begin{tabular}{|c|c|c|c|c|c|c|c|}
$N$ & $\beta$ & $a$ [fm] & $a^{-1}$ [GeV] & $L$ [fm] & Therm. & Sweep & Conf. \cr
\hline
4   & 38.7    & 0.089    & 2.2            & 0.36     & 140    & 14    & 1045  \cr
\hline
4   & 10      & 0.18     & 1.1            & 0.72     & 140    & 14    & 1045  \cr
\hline
10  & 38.7    & 0.089    & 2.2            & 0.89     & 200    & 20    & 1589  \cr
\hline
16  & 38.7    & 0.089    & 2.2            & 1.4      & 260    & 26    & 1017  \cr
\hline
10  & 10      & 0.18     & 1.1            & 1.8      & 200    & 20    & 2365  \cr
\hline
20  & 38.7    & 0.089    & 2.2            & 1.8      & 300    & 30    & 1094  \cr
\hline
26  & 38.7    & 0.089    & 2.2            & 2.3      & 360    & 36    & 1015  \cr
\hline
16  & 10      & 0.18     & 1.1            & 2.9      & 260    & 26    & 1017  \cr
\hline
34  & 38.7    & 0.089    & 2.2            & 3.0      & 440    & 44    & 1038  \cr
\hline
20  & 10      & 0.18     & 1.1            & 3.6      & 300    & 30    & 1041  \cr
\hline
44  & 38.7    & 0.089    & 2.2            & 3.9      & 540    & 54    & 1060  \cr
\hline
26  & 10      & 0.18     & 1.1            & 4.7      & 360    & 36    & 1009  \cr
\hline
56  & 38.7    & 0.089    & 2.2            & 5.0      & 660    & 66    & 1054  \cr
\hline
34  & 10      & 0.18     & 1.1            & 6.1      & 440    & 44    & 1038  \cr
\hline
68  & 38.7    & 0.089    & 2.2            & 6.1      & 680    & 68    & 1011  \cr
\hline
80  & 38.7    & 0.089    & 2.2            & 7.1      & 900    & 90    & 1113  \cr
\hline
44  & 10      & 0.18     & 1.1            & 7.9      & 540    & 54    & 1059  \cr
\hline
104 & 38.7    & 0.089    & 2.2            & 9.3      & 1140   & 114   & 1045  \cr
\hline
56  & 10      & 0.18     & 1.1            & 10       & 660    & 66    & 6564  \cr
\hline
112 & 38.7    & 0.089    & 2.2            & 10       & 1220   & 122   & 3524  \cr
\hline
68  & 10      & 0.18     & 1.1            & 12       & 780    & 78    & 1011  \cr
\hline
136 & 38.7    & 0.089    & 2.2            & 12       & 1460   & 146   & 2179  \cr
\hline
150 & 38.7    & 0.089    & 2.2            & 13       & 1600   & 160   & 1066  \cr
\hline
80  & 10      & 0.18     & 1.1            & 14       & 900    & 90    & 1048  \cr
\hline
166 & 38.7    & 0.089    & 2.2            & 15       & 1760   & 176   & 916   \cr
\hline
184 & 38.7    & 0.089    & 2.2            & 16       & 1940   & 194   & 130   \cr
\hline
104 & 10      & 0.18     & 1.1            & 19       & 1140   & 114   & 1032  \cr
\hline
112 & 10      & 0.18     & 1.1            & 20       & 1220   & 122   & 2250  \cr
\hline
136 & 10      & 0.18     & 1.1            & 25       & 1460   & 146   & 1981  \cr
\hline
150 & 10      & 0.18     & 1.1            & 27       & 1600   & 160   & 2238  \cr
\hline
166 & 10      & 0.18     & 1.1            & 30       & 1760   & 176   & 1286  \cr
\hline
184 & 10      & 0.18     & 1.1            & 33       & 1940   & 194   & 201   \cr
\hline
\end{tabular}
\end{ruledtabular}
\end{table}

\begin{table}
\caption{\label{tmlg3}
Number of configurations considered in the numerical simulations for the results in minimal Landau gauge of section \ref{sgauge} in three dimensions. The value of the lattice spacing $a$ has been determined by interpolation from Ref.\ \cite{Cucchieri:2003di}. To assign units the standard value of (440 MeV)$^2$ has been assigned to the string tension. For further details, see table \ref{tmlg2}.}
\begin{ruledtabular}
\vspace{1mm}
\begin{tabular}{|c|c|c|c|c|c|c|c|}
$N$ & $\beta$ & $a$ [fm] & $a^{-1}$ [GeV] & $L$ [fm] & Therm. & Sweep & Conf. \cr
\hline
4   & 7.09    & 0.090    & 2.2            & 0.36     & 240    & 24    & 1070  \cr
\hline
4   & 4.24    & 0.17     & 1.1            & 0.68     & 240    & 24    & 1070  \cr
\hline
8   & 7.09    & 0.090    & 2.2            & 0.72     & 280    & 28    & 1133  \cr
\hline
12  & 7.09    & 0.090    & 2.2            & 1.1      & 320    & 32    & 1070  \cr
\hline
8   & 4.24    & 0.17     & 1.1            & 1.4      & 280    & 28    & 1070  \cr
\hline
16  & 7.09    & 0.090    & 2.2            & 1.4      & 360    & 36    & 1144  \cr
\hline
20  & 7.09    & 0.090    & 2.2            & 1.8      & 400    & 40    & 1100  \cr
\hline
12  & 4.24    & 0.17     & 1.1            & 2.0      & 320    & 32    & 1128  \cr
\hline
24  & 7.09    & 0.090    & 2.2            & 2.2      & 440    & 44    & 1084  \cr
\hline
16  & 4.24    & 0.17     & 1.1            & 2.7      & 360    & 36    & 1048  \cr
\hline
32  & 7.09    & 0.090    & 2.2            & 2.9      & 520    & 52    & 960   \cr
\hline
20  & 4.24    & 0.17     & 1.1            & 3.4      & 400    & 40    & 1217  \cr
\hline
40  & 7.09    & 0.090    & 2.2            & 3.6      & 600    & 60    & 1065  \cr
\hline
24  & 4.24    & 0.17     & 1.1            & 4.1      & 440    & 44    & 1238  \cr
\hline
48  & 7.09    & 0.090    & 2.2            & 4.3      & 680    & 68    & 1095  \cr
\hline
56  & 7.09    & 0.090    & 2.2            & 5.0      & 760    & 76    & 1078  \cr
\hline
32  & 4.24    & 0.17     & 1.1            & 5.4      & 520    & 52    & 899   \cr
\hline
64  & 7.09    & 0.090    & 2.2            & 5.8      & 840    & 84    & 1062  \cr
\hline
40  & 4.24    & 0.17     & 1.1            & 6.8      & 600    & 60    & 1040  \cr
\hline
48  & 4.24    & 0.17     & 1.1            & 8.2      & 680    & 68    & 1047  \cr
\hline
56  & 4.24    & 0.17     & 1.1            & 9.5      & 760    & 76    & 1053  \cr
\hline
64  & 4.24    & 0.17     & 1.1            & 11       & 840    & 84    & 1033  \cr
\hline
\end{tabular}
\end{ruledtabular}
\end{table}

A second set is obtained and fixed to the minimal Landau gauge using adaptive stochastic overrelaxation \cite{Cucchieri:2006tf}. In this case, a random gauge transformation is used to initialize the iterative procedure. The results from these configurations are used for comparison between the minimal Landau gauge and the absolute Landau gauge in section \ref{sgauge}. The details of these configurations are listed in table \ref{tmlg2} in two dimensions and in table \ref{tmlg3} in three dimensions. The number of configurations was tuned to obtain the exponent of the ghost propagator with a statistical accuracy of better than 10\% at the 1$\sigma$-level.

\begin{table*}
\caption{\label{talg2}
Number of configurations considered in the numerical simulations for the results in absolute Landau gauge of section \ref{sgauge} in two dimensions. For details see table \ref{tmlg2}. The additional parameters $N_w$, $N_p$, $P_p$, $N_c$, $N_m$ are explained in appendix \ref{sabslan}. $N_g$ is the average number of generations, $N_x$ the maximum number of generations, and $N_\mathrm{min}$ , $N_f$, and $N_\mathrm{max}$ are the minimum number, the average number per configuration, and the maximally needed number of gauge-fixings, respectively.}
\begin{ruledtabular}
\vspace{1mm}
\begin{tabular}{|c|c|c|c|c|c|c|c|c|c|c|c|c|c|c|c|c|c|}
$N$ & $\beta$ & $a$ [fm] & $a^{-1}$ [GeV] & $L$ [fm] & Therm. & Sweep & Conf. & $N_W$ & $N_p$ & $P_p$ & $N_c$ & $N_m$ & $N_g$   & $N_x$ & $N_\mathrm{min}$ & $N_f$    & $N_\mathrm{max}$ \cr
\hline
4   & 38.7    & 0.089    & 2.2            & 0.36     & 140    & 14    & 1027  & 8     & 1     & 0.16  & 1     & 2     & 2.36(2) & 6     & 12               & 13.44(8) & 28               \cr
\hline
4   & 10      & 0.18     & 1.1            & 0.72     & 140    & 14    & 1027  & 8     & 1     & 0.16  & 1     & 2     & 2.76(3) & 6     & 12               & 15.0(1)  & 28               \cr
\hline
10  & 38.7    & 0.089    & 2.2            & 0.89     & 200    & 20    & 1208  & 10    & 1     & 1.0   & 2     & 2     & 2.55(2) & 9     & 15               & 17.8(1)  & 50               \cr
\hline
16  & 38.7    & 0.089    & 2.2            & 1.4      & 260    & 26    & 1049  & 12    & 1     & 2.6   & 2     & 3     & 2.58(2) & 7     & 18               & 21.5(1)  & 48               \cr
\hline
10  & 10      & 0.18     & 1.1            & 1.8      & 200    & 20    & 1027  & 10    & 1     & 1.0   & 2     & 2     & 2.65(3) & 8     & 15               & 18.3(2)  & 45               \cr
\hline
20  & 38.7    & 0.089    & 2.2            & 1.8      & 300    & 30    & 1041  & 14    & 2     & 4.0   & 2     & 3     & 2.75(3) & 7     & 21               & 26.3(2)  & 56               \cr
\hline
26  & 38.7    & 0.089    & 2.2            & 2.3      & 360    & 36    & 1130  & 16    & 2     & 6.8   & 3     & 3     & 2.53(2) & 8     & 24               & 28.2(2)  & 72               \cr
\hline
16  & 10      & 0.18     & 1.1            & 2.9      & 260    & 26    & 1037  & 12    & 1     & 2.6   & 2     & 3     & 2.53(2) & 6     & 18               & 21.2(2)  & 42               \cr
\hline
34  & 38.7    & 0.089    & 2.2            & 3.0      & 440    & 44    & 1032  & 18    & 2     & 12    & 3     & 4     & 2.71(3) & 7     & 27               & 33.4(8)  & 70               \cr
\hline
20  & 10      & 0.18     & 1.1            & 3.6      & 300    & 30    & 1119  & 14    & 2     & 4.0   & 2     & 3     & 2.63(2) & 7     & 21               & 25.4(2)  & 56               \cr
\hline
44  & 38.7    & 0.089    & 2.2            & 3.9      & 540    & 54    & 1061  & 20    & 3     & 19    & 4     & 3     & 2.59(3) & 7     & 30               & 35.9(3)  & 80               \cr
\hline
26  & 10      & 0.18     & 1.1            & 4.7      & 360    & 36    & 1065  & 16    & 2     & 6.8   & 3     & 3     & 2.59(3) & 6     & 24               & 28.7(2)  & 56               \cr
\hline
56  & 38.7    & 0.089    & 2.2            & 5.0      & 660    & 66    & 1045  & 24    & 3     & 31    & 4     & 5     & 2.63(3) & 7     & 36               & 43.6(4)  & 96               \cr
\hline
34  & 10      & 0.18     & 1.1            & 6.1      & 440    & 44    & 1032  & 18    & 2     & 12    & 3     & 4     & 2.55(3) & 7     & 27               & 32.0(3)  & 70               \cr
\hline
68  & 38.7    & 0.089    & 2.2            & 6.1      & 680    & 68    & 1130  & 28    & 4     & 46    & 5     & 5     & 2.47(2) & 6     & 42               & 48.6(3)  & 98               \cr
\hline
80  & 38.7    & 0.089    & 2.2            & 7.1      & 900    & 90    & 1084  & 32    & 4     & 64    & 6     & 6     & 2.52(2) & 6     & 48               & 56.3(3)  & 112              \cr
\hline
44  & 10      & 0.18     & 1.1            & 7.9      & 540    & 54    & 1006  & 20    & 3     & 19    & 4     & 3     & 2.46(2) & 6     & 30               & 34.6(2)  & 70               \cr
\hline
104 & 38.7    & 0.089    & 2.2            & 9.3      & 1140   & 114   & 2078  & 36    & 5     & 108   & 7     & 6     & 2.50(2) & 7     & 54               & 63.0(4)  & 144              \cr
\hline
56  & 10      & 0.18     & 1.1            & 10       & 660    & 66    & 1083  & 24    & 3     & 31    & 4     & 5     & 2.56(2) & 7     & 36               & 42.7(2)  & 96               \cr
\hline
112 & 38.7    & 0.089    & 2.2            & 10       & 1220   & 122   & 1068  & 40    & 6     & 125   & 8     & 6     & 2.50(3) & 7     & 60               & 70.0(6)  & 160              \cr
\hline
68  & 10      & 0.18     & 1.1            & 12       & 780    & 78    & 1014  & 28    & 4     & 46    & 5     & 5     & 2.47(2) & 6     & 42               & 48.6(3)  & 98               \cr
\hline
136 & 38.7    & 0.089    & 2.2            & 12       & 1460   & 146   & 1028  & 46    & 6     & 185   & 9     & 8     & 2.49(2) & 7     & 69               & 80.3(5)  & 184              \cr
\hline
150 & 38.7    & 0.089    & 2.2            & 13       & 1600   & 160   & 459   & 52    & 7     & 225   & 10    & 9     & 2.49(3) & 6     & 78               & 90.7(8)  & 182              \cr
\hline
80  & 10      & 0.18     & 1.1            & 14       & 900    & 90    & 2136  & 32    & 4     & 64    & 6     & 6     & 2.45(2) & 8     & 48               & 55.2(3)  & 144              \cr
\hline
166 & 38.7    & 0.089    & 2.2            & 15       & 1760   & 176   & 326   & 58    & 8     & 276   & 11    & 10    & 2.52(5) & 6     & 87               & 102(1)   & 203              \cr
\hline
184 & 38.7    & 0.089    & 2.2            & 16       & 1940   & 194   & 66    & 64    & 9     & 339   & 12    & 11    & 2.45(9) & 5     & 96               & 110(3)   & 192              \cr
\hline
104 & 10      & 0.18     & 1.1            & 19       & 1140   & 114   & 1096  & 36    & 5     & 108   & 7     & 6     & 2.44(2) & 6     & 54               & 61.9(4)  & 126              \cr
\hline
112 & 10      & 0.18     & 1.1            & 20       & 1220   & 122   & 2105  & 40    & 6     & 125   & 8     & 6     & 2.45(2) & 10    & 60               & 69.0(4)  & 220              \cr
\hline
136 & 10      & 0.18     & 1.1            & 24       & 1460   & 146   & 1414  & 46    & 6     & 185   & 9     & 8     & 2.46(2) & 7     & 69               & 79.6(5)  & 184              \cr
\hline
150 & 10      & 0.18     & 1.1            & 27       & 1600   & 160   & 748   & 52    & 7     & 225   & 10    & 9     & 2.51(3) & 7     & 78               & 91.3(8)  & 208              \cr
\hline
166 & 10      & 0.18     & 1.1            & 30       & 1760   & 176   & 267   & 58    & 8     & 276   & 11    & 10    & 2.52(6) & 6     & 87               & 102(2)   & 203              \cr
\hline
184 & 10      & 0.18     & 1.1            & 33       & 1940   & 194   & 53    & 64    & 9     & 339   & 12    & 11    & 2.36(8) & 4     & 96               & 108(3)   & 160              \cr
\hline
\end{tabular}
\end{ruledtabular}
\end{table*}

\begin{table*}
\caption{\label{talg3}
Number of configurations considered in the numerical simulations for the results in absolute Landau gauge of section \ref{sgauge} in three dimensions. See table \ref{tmlg3} and \ref{talg2} as well as appendix \ref{sabslan} for details.}
\begin{ruledtabular}
\vspace{1mm}
\begin{tabular}{|c|c|c|c|c|c|c|c|c|c|c|c|c|c|c|c|c|c|}
$N$ & $\beta$ & $a$ [fm] & $a^{-1}$ [GeV] & $L$ [fm] & Therm. & Sweep & Conf. & $N_W$ & $N_p$ & $P_p$ & $N_c$ & $N_m$ & $N_g$   & $N_x$ & $N_\mathrm{min}$ & $N_f$    & $N_\mathrm{max}$ \cr
\hline
4   & 7.09    & 0.090    & 2.2            & 0.36     & 240    & 24    & 1092  & 8     & 1     & 0.64  & 1     & 2     & 2.70(2) & 6     & 12               & 14.8(2)  & 28               \cr
\hline
4   & 4.24    & 0.17     & 1.1            & 0.68     & 240    & 24    & 1270  & 8     & 1     & 0.64  & 1     & 2     & 2.67(2) & 7     & 12               & 14.7(2)  & 32               \cr
\hline
8   & 7.09    & 0.090    & 2.2            & 0.72     & 280    & 28    & 1115  & 10    & 1     & 5.1   & 2     & 2     & 2.69(3) & 9     & 15               & 18.5(2)  & 50               \cr
\hline
12  & 7.09    & 0.090    & 2.2            & 1.1      & 320    & 32    & 1115  & 12    & 1     & 17    & 2     & 3     & 2.68(3) & 7     & 18               & 22.1(2)  & 48               \cr
\hline
8   & 4.24    & 0.17     & 1.1            & 1.4      & 280    & 28    & 1196  & 10    & 1     & 5.1   & 2     & 2     & 2.73(3) & 7     & 15               & 18.7(2)  & 40               \cr
\hline
16  & 7.09    & 0.090    & 2.2            & 1.4      & 360    & 36    & 1106  & 14    & 2     & 41    & 2     & 3     & 2.62(3) & 8     & 21               & 25.3(2)  & 63               \cr
\hline
20  & 7.09    & 0.090    & 2.2            & 1.8      & 400    & 40    & 1144  & 16    & 2     & 80    & 3     & 3     & 2.64(2) & 8     & 24               & 29.1(2)  & 72               \cr
\hline
12  & 4.24    & 0.17     & 1.1            & 2.0      & 320    & 32    & 1109  & 12    & 1     & 17    & 2     & 3     & 2.54(2) & 6     & 18               & 21.2(1)  & 42               \cr
\hline
24  & 7.09    & 0.090    & 2.2            & 2.2      & 440    & 44    & 1093  & 18    & 2     & 138   & 3     & 4     & 2.56(3) & 7     & 27               & 32.0(3)  & 72               \cr
\hline
16  & 4.24    & 0.17     & 1.1            & 2.7      & 360    & 36    & 1041  & 14    & 2     & 41    & 2     & 3     & 2.54(3) & 8     & 21               & 24.8(2)  & 63               \cr
\hline
32  & 7.09    & 0.090    & 2.2            & 2.9      & 520    & 52    & 1025  & 20    & 3     & 328   & 4     & 3     & 2.68(3) & 8     & 30               & 36.8(3)  & 90               \cr
\hline
20  & 4.24    & 0.17     & 1.1            & 3.4      & 400    & 40    & 1302  & 16    & 2     & 80    & 3     & 3     & 2.60(3) & 8     & 24               & 28.8(2)  & 72               \cr
\hline
40  & 7.09    & 0.090    & 2.2            & 3.6      & 600    & 60    & 1005  & 24    & 3     & 640   & 4     & 5     & 2.73(3) & 8     & 36               & 44.8(4)  & 108              \cr
\hline
24  & 4.24    & 0.17     & 1.1            & 4.1      & 440    & 44    & 1095  & 18    & 2     & 138   & 3     & 4     & 2.57(3) & 7     & 27               & 32.1(3)  & 72               \cr
\hline
48  & 7.09    & 0.090    & 2.2            & 4.3      & 680    & 68    & 906   & 28    & 4     & 1106  & 5     & 5     & 2.84(3) & 8     & 42               & 53.8(4)  & 126              \cr
\hline
56  & 7.09    & 0.090    & 2.2            & 5.0      & 760    & 76    & 657   & 32    & 4     & 1756  & 6     & 6     & 2.83(4) & 8     & 48               & 61.3(6)  & 144              \cr
\hline
32  & 4.24    & 0.17     & 1.1            & 5.4      & 520    & 52    & 1072  & 20    & 3     & 328   & 4     & 3     & 2.71(3) & 9     & 30               & 37.1(3)  & 100              \cr
\hline
64  & 7.09    & 0.090    & 2.2            & 5.8      & 840    & 84    & 167   & 36    & 5     & 2621  & 7     & 6     & 3.02(7) & 6     & 54               & 72(1)    & 126              \cr
\hline
40  & 4.24    & 0.17     & 1.1            & 6.8      & 600    & 60    & 1026  & 24    & 3     & 640   & 4     & 5     & 2.72(3) & 7     & 36               & 44.6(4)  & 96               \cr
\hline
48  & 4.24    & 0.17     & 1.1            & 8.2      & 680    & 68    & 929   & 28    & 4     & 1106  & 5     & 5     & 2.86(3) & 8     & 42               & 54.0(4)  & 126              \cr
\hline
56  & 4.24    & 0.17     & 1.1            & 9.5      & 760    & 76    & 522   & 32    & 4     & 1756  & 6     & 6     & 2.87(4) & 7     & 48               & 61.9(6)  & 128              \cr
\hline
64  & 4.24    & 0.17     & 1.1            & 11       & 840    & 84    & 291   & 36    & 5     & 2621  & 7     & 6     & 3.07(6) & 7     & 54               & 73(1)    & 144              \cr
\hline
\end{tabular}
\end{ruledtabular}
\end{table*}

\begin{table*}
\caption{\label{talg4}
Number of configurations considered in the numerical simulations for the results in absolute Landau gauge of section \ref{sgauge} in four dimensions. See table \ref{tmlg3} and \ref{talg2} as well as appendix \ref{sabslan} for details. The scale has been fixed according to \cite{Bali}.}
\begin{ruledtabular}
\vspace{1mm}
\begin{tabular}{|c|c|c|c|c|c|c|c|c|c|c|c|c|c|c|c|c|c|}
$N$ & $\beta$ & $a$ [fm] & $a^{-1}$ [GeV] & $L$ [fm] & Therm. & Sweep & Conf. & $N_W$ & $N_p$ & $P_p$ & $N_c$ & $N_m$ & $N_g$   & $N_x$ & $N_\mathrm{min}$ & $N_f$    & $N_\mathrm{max}$ \cr
\hline
4   & 2.45    & 0.099    & 2.0            & 0.40     & 340    & 34    & 1020  & 8     & 1     & 2.6   & 1     & 2     & 2.63(3) & 7     & 12               & 14.5(1)  & 32               \cr
\hline
6   & 2.45    & 0.099    & 2.0            & 0.59     & 360    & 36    & 1191  & 10    & 1     & 13    & 2     & 2     & 2.60(3) & 7     & 15               & 18.0(2)  & 40               \cr
\hline
8   & 2.45    & 0.099    & 2.0            & 0.79     & 380    & 38    & 1013  & 12    & 1     & 41    & 2     & 3     & 2.59(2) & 8     & 18               & 21.5(1)  & 56               \cr 
\hline
4   & 2.2     & 0.21     & 0.94           & 0.84     & 340    & 34    & 1132  & 8     & 1     & 2.6   & 1     & 2     & 2.63(2) & 7     & 12               & 14.5(1)  & 32               \cr
\hline
12  & 2.45    & 0.099    & 2.0            & 1.2      & 420    & 42    & 1091  & 14    & 2     & 207   & 2     & 3     & 2.64(3) & 9     & 21               & 25.5(2)  & 70               \cr
\hline
6   & 2.2     & 0.21     & 0.94           & 1.3      & 360    & 36    & 1191  & 10    & 1     & 13    & 2     & 2     & 2.57(3) & 8     & 15               & 17.9(2)  & 45               \cr
\hline
16  & 2.45    & 0.099    & 2.0            & 1.6      & 460    & 46    & 652   & 16    & 2     & 655   & 3     & 3     & 2.65(4) & 8     & 24               & 29.2(3)  & 72               \cr
\hline
8   & 2.2     & 0.21     & 0.94           & 1.7      & 380    & 38    & 1013  & 12    & 1     & 41    & 2     & 3     & 2.56(3) & 6     & 18               & 21.4(2)  & 56               \cr
\hline
20  & 2.45    & 0.099    & 2.0            & 2.0      & 500    & 50    & 148   & 18    & 2     & 1600  & 3     & 4     & 2.72(9) & 10    & 27               & 33.5(7)  & 100              \cr
\hline
12  & 2.2     & 0.099    & 0.94           & 2.5      & 420    & 42    & 1081  & 14    & 2     & 207   & 2     & 3     & 2.65(3) & 7     & 21               & 25.6(2)  & 56               \cr
\hline
16  & 2.2     & 0.21     & 0.94           & 3.4      & 460    & 46    & 277   & 16    & 2     & 655   & 3     & 3     & 2.78(6) & 8     & 24               & 30.2(5)  & 72               \cr 
\hline
20  & 2.2     & 0.21     & 0.94           & 4.2      & 500    & 50    & 59    & 18    & 2     & 1600  & 3     & 4     & 2.7(1)  & 5     & 27               & 33.3(8)  & 54               \cr 
\hline
\end{tabular}
\end{ruledtabular}
\end{table*}

The third set contains the configurations which have been fixed to the absolute Landau gauge. Details on these can be found for two dimensions in table \ref{talg2}, in table \ref{talg3} for three dimensions, and in table \ref{talg4} for four dimensions. Note that in this case it was not always possible to achieve the desired statistical accuracy, as it took, for the largest volume, up to one CPU-week on a current standard processor to obtain one gauge-fixed configuration. The final statistical accuracy achieved was the best obtainable with the available resources. Additional information concerning the gauge-fixing procedure is provided in the tables \ref{talg2}-\ref{talg4}. The description of these can be found in the next appendix \ref{sabslan} on the gauge-fixing algorithm. 

\section{Fixing to the absolute Landau gauge}\label{sabslan}

Fixing to the absolute Landau gauge as specified in section \ref{slandau} is a hard problem. Actually it is NP-hard, similar to finding the ground state of a spin glass Hamiltonian, to the best of current knowledge. Therefore, there is no hope to really find the absolute minimum of \pref{fumod}. In particular, confirming to have found the absolute minimum of \pref{fumod} is in itself a problem at least as hard as the original problem of finding the minimum. Therefore, the best attempt, which is pursued here, is to improve the condition \pref{propmin}. That this at least resembles the absolute Landau gauge rests on the assumption that the evolution of all (relevant) quantities with the value of \pref{fumod} is smooth, and the limit itself is also not discontinuous. Hence, by getting closer and closer to the absolute minimum the propagators should become more and more like the ones in the correct absolute Landau gauge. On the smallest lattices, there is a reasonable hope to have found the absolute minimum in most of the cases. There, no discontinuous behavior has been found, giving at least some justification for this assumption.

It is then necessary to design an algorithm which aims at improving the condition \pref{fumod}. This has often been pursued using multiple restart algorithms \cite{Cucchieri:1997ns,mmp4,Cucchieri:1997dx}, stimulated annealing \cite{l72} and other approaches \cite{Bogolubsky,Silva:2004bv,Oliveira:2003wa,Yamaguchi:1999hp}. However, the problem of gauge-fixing is, as has been discussed in section \ref{sgf}, highly configuration-dependent. It is therefore desirable to use an algorithm which is able to adapt to a given configuration. One possibility are genetic (or evolutionary) algorithms \cite{genetic}. This type of algorithms has been used previously for gauge-fixing in lattice gauge theory \cite{Silva:2004bv,Oliveira:2003wa,Yamaguchi:1999hp}. Here, a different flavor of such an algorithm is constructed and used throughout section \ref{sgauge}.

The following algorithm has been performed for any given configuration ${\cal C}$:

\begin{itemize}

\item[1)] Generate an initial population of $N_W$ gauge transformations. They are obtained using an adaptive stochastic overrelaxation algorithm from once the identity transformation and $N_W-1$ random gauge transformations, and all fix ${\cal C}$ to the Landau gauge\footnote{Landau gauge refers here to the (local) minimum of \pref{fumod} which is reached by using adaptive stochastic overrelaxation \cite{Cucchieri:2006tf}.}

\item[2)] Determine the fitness of each member of the population by calculating \pref{fumod} for it

\item[3)] Remove the $N_W/2$ elements with the lowest fitness function, obtaining the parent generation ${\cal P}$. Note that all digits of the fitness function are always considered, i.\ e., even two members with a fitness only differing by machine precision fluctuations are not considered equal. Then construct a new generation by the following operations

\begin{itemize}

\item[a)] Generate 0.15$N_W=N_p$ new members by point mutation\footnote{Non-integer numbers are always rounded mathematically, but set to one if zero.}. For this, select one member from ${\cal P}$ randomly. Then draw a random number $R$ according to a Gaussian distribution with a given width $P_p$. Take the absolute value, and if the number is smaller than 1 set it to one. Replace at $R$ randomly chosen lattice points $x$ the local gauge transformation matrix $g(x)$ by a randomly chosen SU(2) matrix. Permit multiple replacements. However, $P_p$ should be chosen sufficiently small. Here, it was selected to be 1\% of the number of lattice sites. Use the so generated initial gauge transformation as seeds to obtain once again a gauge transformation of ${\cal C}$ to Landau gauge

\item[b)] Generate 0.2$N_W=N_c$ new members by crossing. For this purpose select two different members from ${\cal P}$ randomly. Then construct a new member by mixing these two parent members. For this, select at each lattice site randomly from which parent to take the matrix $g(x)$. Use this new member as a seed to obtain a gauge transformation of ${\cal C}$ to Landau gauge

\item[c)] Add an additional number $N_m$ of new members such that the population again consists of $N_W$ members. These are again randomly initialized gauge transformations, which are then used to construct gauge transformations of ${\cal C}$ to Landau gauge

\end{itemize}

\item[4)] Determine the fitness of all members according to \pref{fumod}. If none of the new members are better than those of the set ${\cal P}$, terminate the algorithm and use the fittest member for the gauge transformation to absolute Landau gauge. If a better new member exists, go to point 3 and repeat until the termination criterion is satisfied. A new member is also considered better if it only differs from the best so far in fitness by an amount of the order of the machine precision

\end{itemize}

Note that this algorithm does not guarantee to terminate after a number of generations. However, in practice it always terminates. In most cases, already the second generation did not yield any improvement, but the maximum number of generations $N_x$ has been as large as 10 for the systems investigated here. The average number of generations $N_g$ is always between two and a bit more than three, yielding therefore a total number of performed gauge-fixings between $3/2N_W$ and roughly $2N_W$. Given that the number of Gribov copies grows exponentially with the number of lattice sites, $N_W$ also has to grow with the number of lattice sites. Strict exponential growth is at this time prohibited by computational resources. The values chosen in practice are listed in tables \ref{talg2}-\ref{talg4}.

Note that the numbers $N_W$, $N_p$, $N_c$, $N_m$, and $P_p$ are far from optimized, and result only from experiments on small lattices. This algorithm can and should be improved, if fixing to the absolute Landau gauge turns out to be important for applications. Furthermore, the structure of the algorithm is itself subject to optimization, and may possibly be more efficient if other structures were admitted as well \cite{genetic}. Finally, there exists some knowledge on the structure of Gribov copies \cite{Heinzl:2007cp}. In particular, they exhibit a domain-like structure. Incorporating this prior knowledge in, e.\ g., the crossing of two gauge transformations may furthermore improve the efficiency of the algorithm.

%%%%%%%%%%%%%%%%%%%%%%%%%%%%%%%%%%%%%%%%%%%%%%%%%%%%%%%%%%%%%%%%%%%%%%%%%%%%%%%%%%%%%%%%%%%%%%%%%%%

\end{document}